# Initial Stages of Rejuvenation of Vapor-Deposited Glasses during Isothermal Annealing: Contrast Between Experiment and Simulation

M. E. Tracy*,[1], B. J. Kasting[1], C. Herrero[2,3], L. Berthier[2,4], R. Richert[5], A. Guiseppi-Elie[6], M. D. Ediger[1]

[1] *Department of Chemistry, University of Wisconsin-Madison, 1101 University Ave, Madison, Wisconsin 53706, USA*

[2] *Laboratoire Charles Coulomb (L2C), Université de Montpellier, CNRS, 34095 Montpellier, France*

[3] *Institut Laue-Langevin, 71 Avenue des Martyrs, 38042 Grenoble, France*

[4] *Gulliver, UMR CNRS 7083, ESPCI Paris, PSL Research University, 75005 Paris, France*

[5] *School of Molecular Sciences, Arizona State University, Tempe, Arizona 85287, USA*

[6] *Department of Electrical and Computer Engineering, Texas A&M University, College Station, Texas 77843, USA*

* corresponding author: megantracy88@gmail.com

**Abstract:**

Physical vapor deposition can prepare organic glasses with high kinetic stability. When heated, these glassy solids slowly transform into the supercooled liquid in a process known as rejuvenation. In this study, we anneal vapor-deposited glasses of methyl-m-toluate (MMT) for six hours at $0.98T_g$ to observe rejuvenation using dielectric spectroscopy. Glasses of moderate stability exhibited partial or full rejuvenation in six hours. For highly stable glasses, prepared at substrate temperatures of $0.85T_g$ and $0.80T_g$, the six-hour annealing time is ~2% of the estimated transformation time, and no change in the onset temperature for the α relaxation process was observed, as expected. Surprisingly, for these highly stable glasses, annealing resulted in significant increases in the storage component of the dielectric susceptibility, without corresponding increases in the loss component. These changes are interpreted to indicate that short-term annealing rejuvenates a high frequency relaxation (e.g., the boson peak) within the stable glass. We compare these results to computer simulations of the rejuvenation of highly stable glasses generated by the swap Monte Carlo algorithm. The *in silico* glasses, in contrast to the experiment, show no evidence of rejuvenation within the stable glass at times shorter than the alpha relaxation process.

I.   Introduction

Glasses have many applications in modern life, well beyond windowpanes and eyeglasses. For example, the layers of organic emitters in organic light emitting diode (OLED) screens found in most modern smart phones are glasses of organic semiconductors[1]. Many applications that require optical transparency, like optical fibers, or compositional flexibility, like organic electronics, also use glassy materials[2-4]. Additionally, there has been much interest in the potential applications of glasses of metal mixtures, whose high elasticity and high strength, as well as their resistance to corrosion, make them intriguing prospects for many advanced technological applications[5-9].

Glassy materials are inherently nonequilibrium and have properties that depend upon their method of preparation. Glasses are typically formed by cooling a liquid to temperatures below its melting point, into a metastable state called the supercooled liquid, then continuing to cool until the molecules or atoms no longer have enough mobility to maintain the liquid's properties (i.e. density and enthalpy). The temperature at which this change occurs is referred to as the glass transition temperature ($T_g$), and below this point the liquid is considered to have formed a glass[2, 10]. Cooling may be conducted at different rates. Cooling more slowly or cooling followed by isothermal annealing both lead to denser glasses with lower enthalpy and improved physical properties such as elastic modulus and thermal stability[11-17]. However, given the range of practical cooling rates and annealing times, only marginal improvements in properties can be attained by this approach[2, 18-21].

In contrast to liquid cooling, physical vapor deposition (PVD) provides a practical approach to access low enthalpy and high stability glass states[22]. This method of forming a glass involves the deposition of molecules out of the gas phase onto a temperature-controlled substrate. When the substrate is held at temperatures somewhat below $T_g$, the enhanced mobility of the molecules at the free surface allows fast equilibration toward the liquid at the substrate temperature[23-30]. For higher deposition temperatures and slower deposition rates, the metastable supercooled liquid seems to be reached, while lower deposition temperatures and faster deposition rates create glasses with high density and low enthalpy that are distinct from the supercooled liquid[26, 31, 32]. A major advantage of the PVD technique is that it forms glasses with low enthalpy and high stability on a timescale of minutes, rather than the many years it would take to reach such low-energy states through liquid-cooling[18, 21, 33]. PVD is thus an advanced materials processing technique that rapidly offers glasses with significant technological advantage.

Stable PVD glasses uniquely allow systematic observation of the process of rejuvenation for samples with low enthalpy and high stability. The primary structural relaxation (also called the α relaxation) in these glasses is shifted to extremely long times, such that it is impossible to directly observe. When stable glasses are heated above $T_g$, the α relaxation process is slow to grow back in, as indicated by an elevated onset temperature ($T_{ons}$) where the glass recovers to the liquid state[33, 34]. The mechanism of this rejuvenation of the primary relaxation process for stable glasses is of particular interest, as it gives insight into the structure and dynamics of these low-energy states. Previous work has shown that these transformations have several unique features. A stable PVD film heated above $T_g$ initiates a front of supercooled liquid that grows through the film from the free surface; surface initiation results from enhanced mobility at the surface relative to the bulk[35-37]. For thick films (or when the front mechanism is suppressed by a capping layer[38]), a bulk mechanism also takes place. This involves nucleation of 'bubbles' of liquid within the rigid matrix of the dense stable glass. These liquid regions then grow with time until the whole film has transformed[37-42]. These rejuvenation features are generic signatures of highly stable glasses (rather than PVD glasses in particular) as they have all been observed in computer simulations that utilize highly equilibrated configurations (not generated by deposition)[39, 43]. Of particular relevance are simulations that use the swap Monte Carlo algorithm[44] to generate ultrastable amorphous packings that rival the stability of experimental PVD glasses[45, 46].

Highly stable PVD glasses also show partial suppression of relaxation processes that occur below $T_g$. Examples include the suppression of the Johari-Goldstein beta relaxation[34, 45, 47] as well as of other intramolecular relaxation processes[42, 48], and also a reduction in the number of quantum tunneling two-level systems[49, 50]. The high frequency boson peak, which is a relaxation process present in all disordered materials[51-55], is also suppressed in stable glasses[56-58]. This process appears as an excess in the vibrational density of states (VDOS) of the glass relative to that of the Debye crystal, and can be observed in the low-temperature heat capacity as well as in measurements like neutron and Brillouin light scattering. The boson peak also appears in dielectric spectra between 100 and 1000 GHz[59]. There are many physical pictures that have been used to explain the origin of the boson peak, ranging from phonon modes[52, 54, 60] to relaxation of clusters of locally preferred structures[51] within the glass, without any strong consensus. Regardless of the origin, it has been observed that increasing the density of the glass, either through increasing the pressure[61, 62] or by decreasing the cooling rate[63] or otherwise forming a more stable glass[64], results in a decrease of the boson peak intensity.

In contrast to previous studies of the rejuvenation of stable PVD glasses above $T_g$, this work seeks to understand rejuvenation below $T_g$, with a focus on the initial stages of rejuvenation. We quantify rejuvenation by examining changes in the dielectric susceptibility for vapor deposited films of methyl-m-toluate (MMT, $T_g = 170$ K) annealed at $0.98T_g$ for up to six hours. While less stable glasses of MMT completely rejuvenate on this timescale, the recovery of the α peak on heating of the most stable films is unaffected by this annealing. Surprisingly, the most stable glasses show an increase in the storage component of dielectric susceptibility during annealing, suggesting that the boson peak (at frequencies above the experimental range) is substantially recovering even though the annealing time is only ~2% of the time required to fully rejuvenate these glasses. This appears to indicate a "softening" of the stable glass long before it transforms into the supercooled liquid, potentially related to the softening reported previously[65].

Given the surprising nature of the experimental observations, we make careful comparisons with swap Monte Carlo computer simulations on a two-dimensional polydisperse systems. In a procedure analogous to the experiments, the rejuvenation of a highly stable glass has been observed after a temperature upjump. Previous work has shown that these *in silico* glasses rejuvenate in a manner similar to the most stable experimental glasses (prepared by PVD)[39, 43]. However, here we find no feature in the computer simulations that can be associated with the experimentally observed increase in the storage component of the dielectric response. That is, the computer simulations show no evidence of softening of the stable glass, prior to its transformation to the supercooled liquid. We present some speculative ideas for this contrast between experiments and computer simulations.

II. Methods

A. Experiments

Liquid MMT of 98% purity was purchased from Alfa Aesar and the material was used as received. Films of MMT were vapor-deposited onto a temperature-controlled substrate within a custom-built vacuum deposition chamber, with typical base pressure lower than $10^{-9}$ torr. Glassy films were prepared using substrate temperatures below Tg, specifically 138.2 K (0.80$T_g$, forming the glass with the highest stability), 143.2 K (0.85$T_g$), 153.2 K (0.90$T_g$) and 161.2 K (0.95$T_g$, forming the glass with the lowest stability). The typical thickness of these films was 5 μm with a typical rate of deposition about 1.8 ± 0.5 nm/s. Variations in deposition rate did not measurably affect the stability of the prepared films. The procedure for determination of the deposition rate, and the thickness of the resulting films, is discussed in the SI. Corresponding liquid-cooled glasses were generated by transforming the stable samples by heating above $T_g$ and then cooling at 5 K/min.

To allow the dielectric measurements to be conducted *in situ*, the film was deposited onto an interdigitated electrode (IDE), IME 1025.3-FD-Pt-U manufactured by ABTECH Scientific, Inc. This IDE has two pairs of platinum fingers that were microlithographically patterned in the lift off method onto a borosilicate glass substrate. Each pair of electrodes consists of a total of 50 interdigitated fingers that are 10 μm wide and spaced 10 μm apart, leading to a geometric capacitance of 0.65 pF. One of these pairs of electrodes was exposed to the vapor deposition and formed a film of deposited material, while the other is enclosed to prevent deposition. The temperature of the substrate was measured by a three-wire RTD and controlled with a heater connected to an Omron temperature controller.

To measure the dielectric susceptibility of the film, a Solartron SI-1260 applies an alternating voltage ($V_{rms}$ = 2.9 V) across the electrodes; the resulting current for each pair of electrodes is then measured using two transimpedance amplifiers[66, 67]. The relationship between current and voltage gives the experimental complex impedance according to $Z^* = \frac{V^*}{I^*}$. This impedance is converted to the dielectric susceptibility ($\chi^* = \chi' - i\chi''$) as described previously[68].

We used two different methods to account for the substantial temperature-dependent contribution to the susceptibility values from the borosilicate substrate of the IDE. For isothermal measurements, a constant-value background-subtraction method was employed. The values for each experiment were taken as the average of the near-constant baseline, as measured for ten minutes at the annealing temperature before film deposition. This constant value was then subtracted from the measured real and imaginary susceptibilities for each frequency, as obtained during annealing. For experiments where temperature was not held constant, a two-channel subtraction method was instead employed, making use of the second, covered pair of electrodes to provide the substrate contribution at each temperature.

Two types of isothermal annealing experiments were performed. The first focused on monitoring the dielectric susceptibility at a pair of frequencies, 20 Hz and either 610 Hz or 20 kHz, during annealing at 167.2 K (0.98$T_g$). These frequencies were chosen to minimize the contribution of noise from the system and surroundings, and are located in the portions of the spectral window where the high frequency wing of the α peak and the low frequency wing of the beta peak overlap (20 Hz), or in the high frequency wing of the beta peak (610 Hz or 20 kHz). The second type of experiment took full spectra ($10^5$ Hz to $10^0$ Hz, with 8 points per decade spaced logarithmically) during the same annealing protocols. This second measurement protocol sacrifices time resolution for more detail on the relaxation processes occurring during the annealing since each spectrum took about 10 minutes to collect.

After isothermal annealing, the glassy film was transformed into the supercooled liquid by heating at 5 K/min. The kinetic stability of the film was assessed using the onset temperature on heating. The onset temperature, which is defined to be representative of the bulk transformation process for these thick PVD films, is determined by fitting a line to the loss component at low temperatures and then finding its intersection with a line fitted to the final approach to the supercooled liquid curve, as shown in Supplemental Figure 3.

The transformation of the corresponding liquid-cooled film, formed by cooling the supercooled liquid at 5 K/min, was observed in the same manner. This allowed us to confirm that our temperature calibration was consistent within 0.5 K and that the film showed no loss of signal due to crystallization. Crystallization was not observed if the film remained below roughly 195 K.

We fit data from these experiments to Kohlrausch-Williams-Watts (KWW) functions in order to parameterize the shape and timescale of changes observed. This function has the form $\frac{(T_{ons}(t)-T_{ons,\infty})}{(T_{ons}(0)-T_{ons,\infty})} = e^{-(t/\tau_{KWW})^\beta}$, where t is time, $T_{ons}$(t) is the onset temperature at time t, $T_{ons}$(0) is the onset temperature of the unannealed sample, $T_{ons,\infty}$ the equilibrium onset temperature at the annealing temperature, β is the stretching exponent describing the deviation from an exponential form, and $\tau_{KWW}$ is the characteristic timescale.

B. Simulations

Following previous work[39], we performed molecular dynamics (MD) simulations on a two-dimensional size-polydisperse mixture composed of soft repulsive spheres. We refer to this publication for details regarding the particle size distribution, and the pairwise potential between particles. We employ reduced units based on the mass $m$ of the particles and the energy scale $\varepsilon$ of the pair potential, to express timescales in units of the microscopic time $\tau_{LJ} = \sqrt{m/\varepsilon}$, where $\sigma$ is the unit length defined as the average particle diameter. We use a system composed of $N = 64000$ particles in a periodic simulation box. We studied the rejuvenation of ultrastable systems prepared using the swap Monte Carlo algorithm. The stability of the system is encoded in the preparation temperature $T_i$ used to create these configurations. Here we used the most stable configurations created at $T_i = 0.035 \sim T_g/2$. Rejuvenation is induced by suddenly heating the system to a higher temperature $T_a = 0.1 \sim 1.4\,T_g$, keeping the pressure constant. Compared to the experiments, the annealing temperature is larger, as rejuvenation at lower $T_a$ is too slow to be observable in our numerical time window.

At this $T_a$ the transformation time is larger than $10^7$, and we can thus easily analyze the behavior of the system at times much before complete rejuvenation, in analogy with the experiments. The dynamics of the system is analyzed using the mean-squared displacements (MSD) of the particles

$$\Delta^2(t_w, t_w + t) = \frac{1}{N}\sum_i^N |r_i(t + t_w) - r_i(t_w)|^2 \qquad (1)$$

where $r_i(t)$ is the position of the particle $i$ at time $t$, and the rejuvenation process starts at $t_w = 0$. By systematically varying $t_w$ and t in the MSD, we can resolve the dynamics of the system over a broad time window, at various stages of the rejuvenation process.

III. Experimental Results

In these experiments, we observed changes in the dielectric response of vapor-deposited glasses of MMT during and after 6 hours of isothermal annealing. Each individual experiment consists of three parts, as shown in Figure 1: the deposition, during which the substrate temperature is constant, controlling the stability of the deposited film; the sub-$T_g$ annealing, conducted at $0.98T_g$ (167.2 K) to allow the film to partly or completely rejuvenate; and the full heating above $T_g$, which reads out the effect of the annealing protocol on the α relaxation process based on the onset temperature. Below, we first discuss the results observed during the annealing step, and then separately results from the transformation step.

The observed dielectric property used here is the complex dielectric susceptibility, $\chi^* = \chi' - i\chi''$. This is similar to the more commonly used dielectric permittivity, $\varepsilon^*$, with the difference being that the real component starts from a value of zero when there is no contribution from a sample, as compared to ε' which starts at a value of 1. $\chi^*$ is more convenient for our discussion of relative changes of the dielectric response with time.

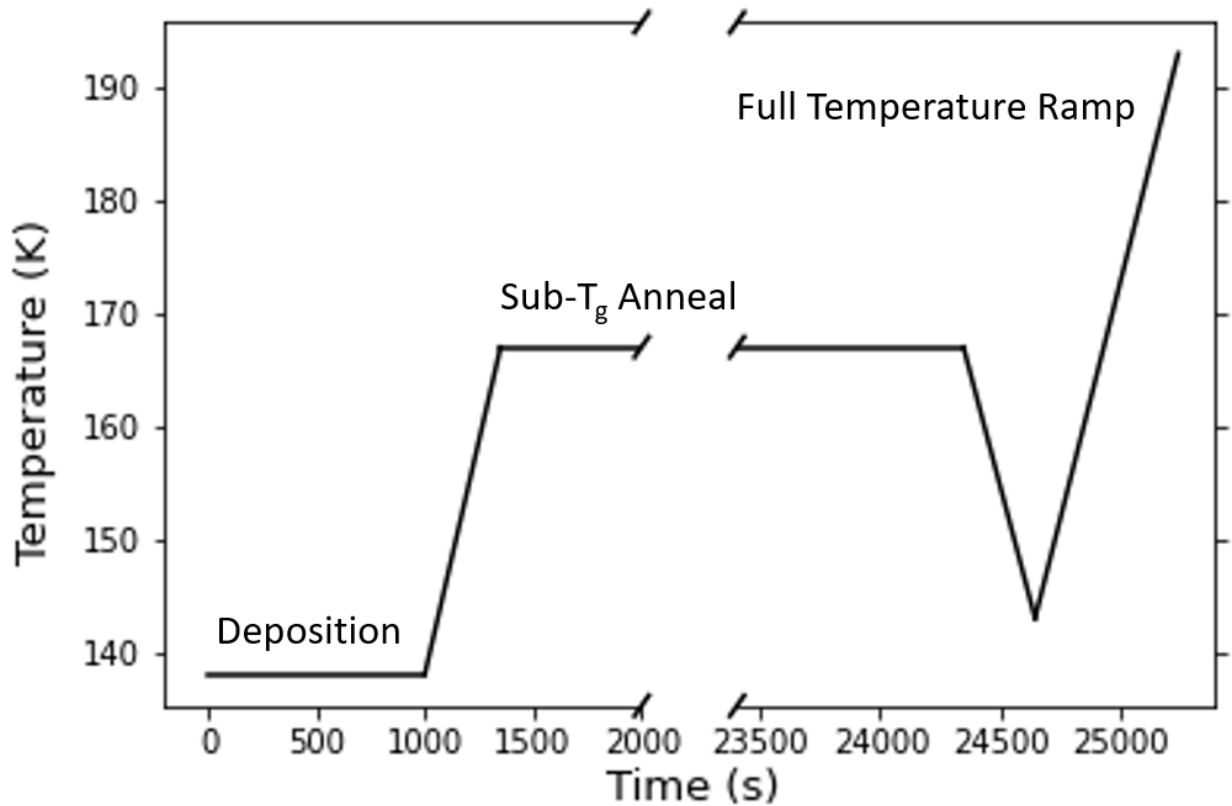

Figure 1: Temperature profile schematic of a typical experiment on a vapor-deposited glass of MMT, highlighting the three important stages of the experiment: deposition at temperatures below $T_g$, annealing at $0.98T_g$, and finally heating above $T_g$ to observe the onset temperature characterizing the return of the film to equilibrium.

A. Isothermal annealing below $T_g$

During annealing at $0.98T_g$, the vapor-deposited glasses of MMT with different stabilities responded qualitatively differently, as reflected in Figure 2. For the somewhat stable films (those deposited at $0.95T_g$), the glasses rejuvenate completely during annealing: their storage and loss increase to a new steady state, as expected for a film that reaches equilibrium at $0.98T_g$. To confirm that the final steady state is the equilibrium state, we aged a liquid-cooled glass to equilibrium at $0.98T_g$. The values at the end of that experiment (shown by the magenta arrows and discussed in more detail in the SI), are in good agreement with the final steady state of the deep red curves in Figure 2. The shape of this relaxation, which fits to a KWW function with exponent $\beta = 1.1$ as shown in Supplemental Figure 7, is consistent with a decrease in density toward equilibrium after a temperature up-jump[69]. As we discuss below, an equilibration time of $\sim 10^4$ s is reasonable for a glass of this stability at this annealing temperature.

Films deposited at $0.90T_g$ show partial rejuvenation during annealing at $0.98T_g$, as seen by an increase in the storage component and the smaller increase in loss component. If this film had reached equilibrium during annealing, the final dielectric observables would have matched the final values attained for the $0.95T_g$ glass.

In contrast to these less stable films, the most stable glasses are those deposited at $0.85T_g$ and $0.80T_g$, which show increases in the real component of dielectric susceptibility only (Figure 2A), with no appreciable change in the loss component (Figure 2B). Specifically, in 6 hours of annealing, the real component of the most stable films increases by almost 20% of the difference between the initial stable glass and the final equilibrium state at $0.98T_g$. While Figure 2 shows dielectric results obtained at 20 Hz, similar results were obtained at two other frequencies (610 Hz and 20 kHz) as shown in Supplemental Figure 8.

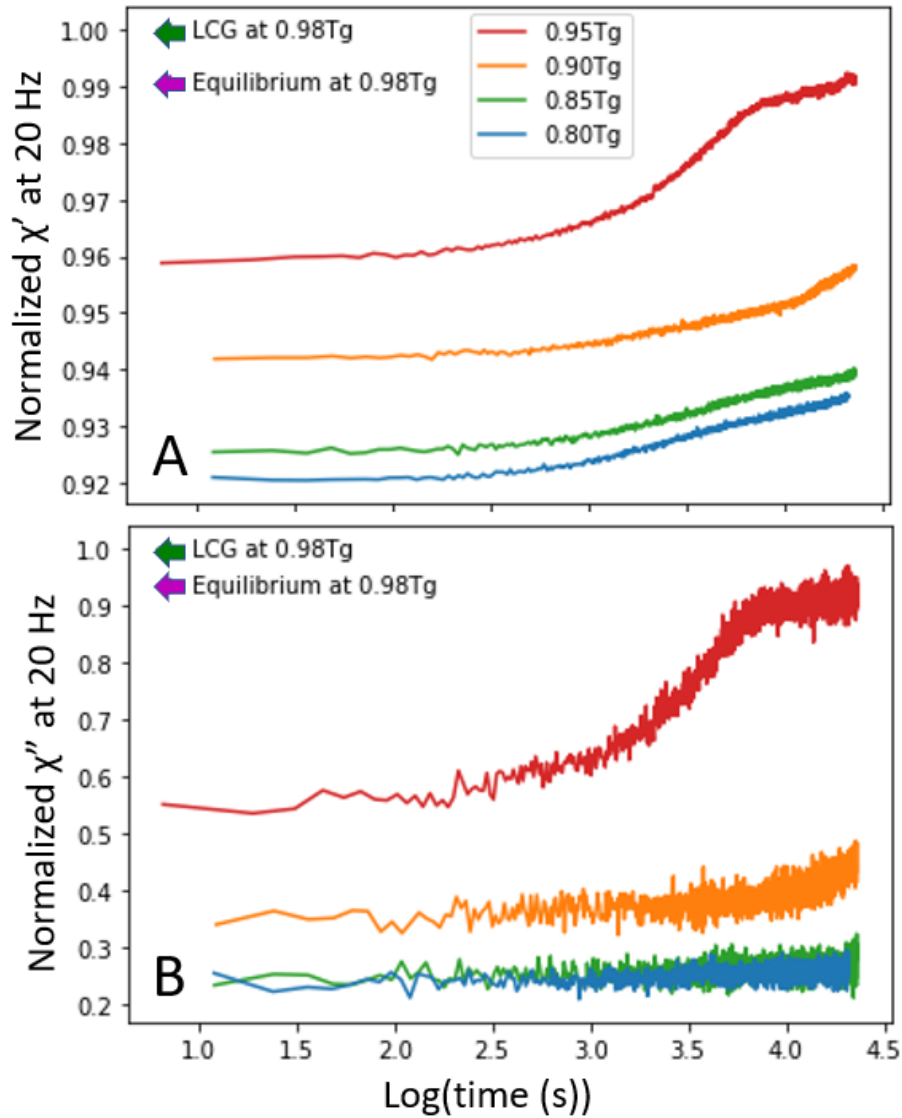

Figure 2: Normalized changes in χ' (storage component, A) and χ" (loss component, B) measured at 20 Hz, as a function of annealing time at $0.98T_g$ for PVD glasses deposited at different substrate temperatures. To account for slight differences in sample thickness, these values are normalized by dividing by the value of the same quantity measured for the corresponding liquid-cooled glass at $0.98T_g$ for each film, defined at 1 as indicated by the green arrows. The value (normalized in the same way) for a film aged to equilibrium at $0.98T_g$ is represented by the magenta arrow in each plot.

The observation that the most stable PVD glasses show an appreciable increase in the storage component without any apparent change in the loss was notable and surprising. To investigate this further, we conducted equivalent experiments on the most stable glasses ($0.80T_g$), taking full dielectric susceptibility spectra during 6 hours of annealing to more fully understand the observed features in the single-frequency experiments. As seen in Figure 3, the increase in the storage component is uniform across the entire spectrum. The data shown in Figure 3 is fully consistent with that shown in Figure 2.

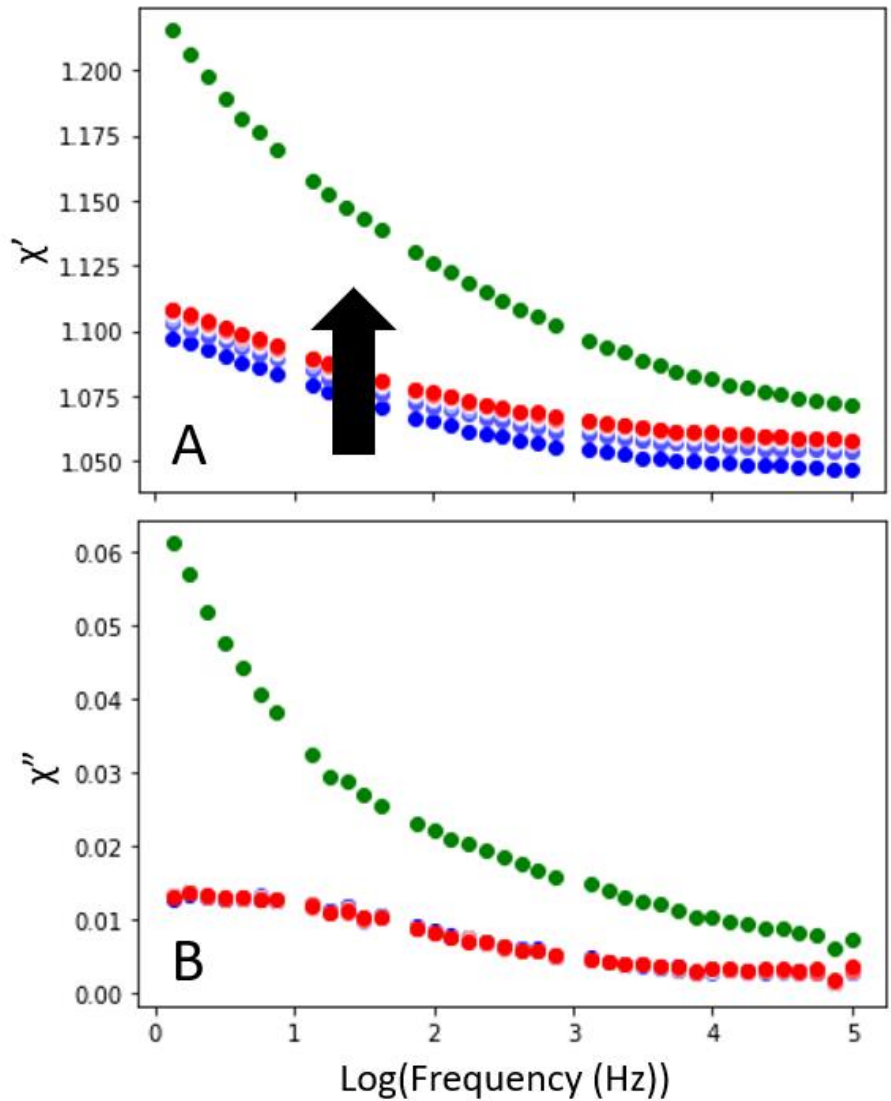

Figure 3: Storage (A) and loss (B) components of a $0.80T_g$ MMT glass during annealing at $0.98T_g$. To decrease noise, the 34 spectra acquired were averaged in groups of 5 (with one group of 4). The earliest spectrum is shown in blue, while the last is shown in red, with the gradient between the two representing the progression of annealing time. Selected frequencies were removed from all spectra due to noise issues. The spectrum measured at the same temperature for the corresponding liquid-cooled glass of this same film are shown in green. The broad peak centered near 10 Hz is the β relaxation process (the amplitude is suppressed in the PVD glass).

We return to the interpretation of these results in the Discussion section.

### B. Temperature ramping into the supercooled liquid

A second way to examine the effect of the annealing on the state of the glass is to measure the onset temperature upon heating after annealing has been completed. An onset temperature which is high is an indicator of high kinetic stability in the glass, and correlates well with a decrease in enthalpy[32, 70, 71]. Here, we transform the deposited films after different annealing times to look for changes in the onset temperature as a second way to quantify the extent of rejuvenation. The temperature ramping experiments revealed a simple trend as a function of annealing time and glass stability.

The films deposited at $0.95T_g$ rejuvenate completely during annealing at $0.98T_g$, as seen in Figure 4A. The as-deposited glass has a low value of $\chi"$ until the onset of the transformation into the supercooled liquid ($T_{ons} \approx 174$ K). The sample completely transforms within a few Kelvin, and joins the equilibrium supercooled liquid response at about 177 K. The large peak near 183 K is the α relaxation process for the supercooled liquid measured at 20 Hz. Annealing at $0.98T_g$ systematically reduces $T_{ons}$, and about 5 ks of annealing completes the transformation, in agreement with Figure 2.

To facilitate later discussion, the evolution of $T_{ons}$ with annealing time is summarized in Figure 5 for MMT glasses with a range of initial stabilities. The journey to equilibrium for the $0.95T_g$ film value occurs on the timescale of a few thousand seconds at $0.98T_g$, as seen in Figure 4A. Since $\tau_\alpha$ at $0.98T_g$ is on the order of 1000 s (based on a Vogel-Fulcher-Tammann (VFT) extrapolation of literature data[72]) and as this is not a particularly stable glass, this transformation time is expected.

Figure 4B presents the evolution of the $0.90T_g$ glasses during annealing at $0.98T_g$. These glasses show only partial rejuvenation over 23 ks, with a gradual decrease of the onset temperature during annealing. This is consistent with the results in Figure 2, which also indicated an only partial rejuvenation. Figure 5 indicates that the onset temperature shifts less than half of the way towards equilibrium over 23 ks.

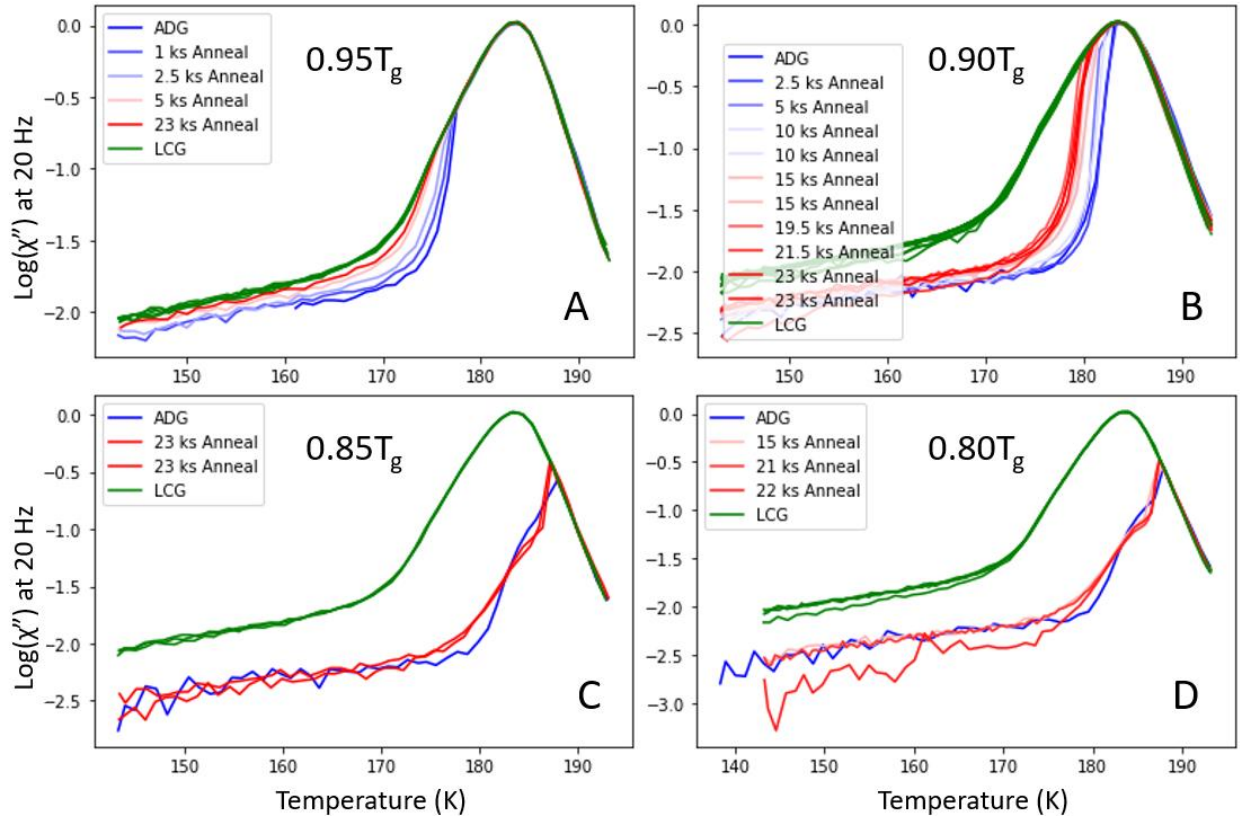

Figure 4: Normalized values of χ" on heating (5 K/min) for MMT glasses deposited at $0.95T_g$ (A), $0.90T_g$ (B), $0.85T_g$ (C), and $0.80T_g$ (D). The unannealed deposited glass is shown in the darkest blue, while the film with the longest annealing time is the darkest red. The corresponding heating curves for the liquid cooled films associated with each deposition is included in green for comparison.

Finally, the most stable MMT glasses (deposited at $0.85T_g$ in Figure 4C, deposited at $0.80T_g$ in Figure 4D) show very little or no change in the onset temperature due to annealing, even at the longest annealing times. These results, summarized in Figure 5, are consistent with the picture that the re-equilibration of the α process in these films has barely begun in 23 ks, or not even started. Based upon measurements over a range of annealing temperatures, we estimate that the time required to equilibrate the $0.80T_g$ sample at $0.98T_g$ is roughly $10^6$ s (see Supplemental Figure 9). The observation that MMT glasses deposited at $0.80T_g$ and $0.85T_g$ have high kinetic stability is consistent with previous experiments[34, 37, 73].

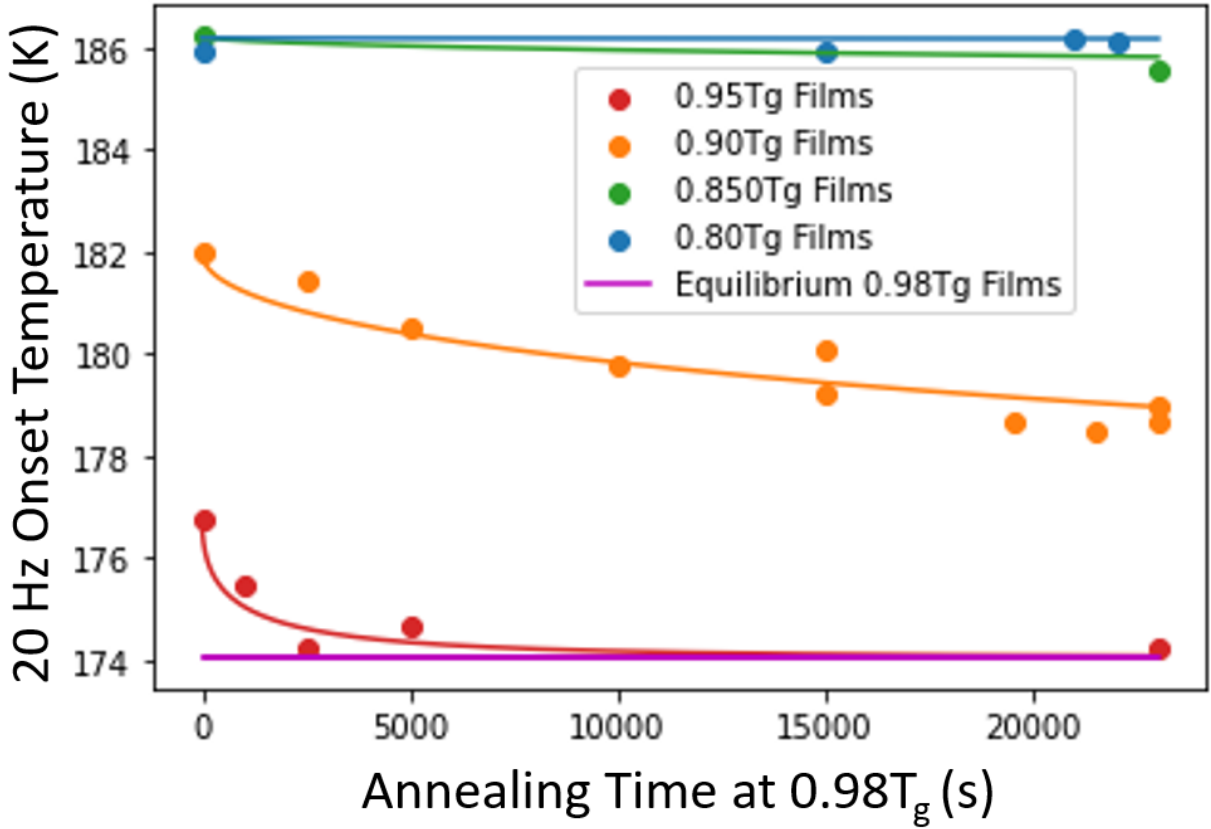

Figure 5: Onset temperature as a function of annealing time at $0.98T_g$ for MMT glasses of different stabilities. The onset temperature of a liquid-cooled glass that had been aged to equilibrium at $0.98T_g$ is represented by the magenta line. For each deposition temperature, the evolution of the onset temperature with annealing time was fitted to a KWW function, shown by the solid lines.

IV.  Discussion and Comparison with Simulations

We have isothermally annealed PVD glasses of MMT at $0.98T_g$ for up to 23 ks. Films thicker than 3 μm were used in these experiments in order to access the bulk dynamics rather than any effects of the free surface, such as the transformation fronts observed in thinner films. We have characterized the rejuvenation of these glasses in two ways: 1) changes in the dielectric susceptibility during annealing (Figures 2 and 3), and 2) changes in the temperature-induced transformation brought about by annealing (Figures 4 and 5). For the two least stable glasses (deposited at $0.95T_g$ and $0.90T_g$), results from the two methods of characterization are easily and consistently interpreted: the sample deposited at $0.95T_g$ completely rejuvenates during 23 ks of annealing, and the sample deposited at $0.90T_g$ partially rejuvenates.

On the other hand, results from the two methods of characterization are not easily understood for the two most stable MMT glasses (deposited at $0.85T_g$ and $0.80T_g$). Results obtained during temperature ramping show essentially no rejuvenation of the α process as a result of annealing at $0.98T_g$, while $\chi'$ recovers approximately 20% during annealing, relative to a fully rejuvenated sample. How does $\chi'$ recover significantly without a significant change in the α process? And how does $\chi''$ remain constant while $\chi'$ is significantly changing?

### A. Recovery of the boson peak

The increase of $\chi'$ without increase of $\chi''$ at the same frequency, as shown for the most stable glasses in Figures 2 and 3, indicates a change in the intensity of some process taking place at frequencies higher than our spectral window. This conclusion follows from the Kramers-Kronig relations, and we illustrate this idea schematically in Figure 6. We have speculatively identified this process as the boson peak, which is universal to all amorphous materials[51-55]. Despite our inability to directly observe the boson peak in this work, it is reasonable to suppose that the boson peak was suppressed in the original stable glass relative to a rejuvenated sample. This idea is supported by previous observations of the boson peak suppression in stable PVD glasses[49, 58], and is further consistent with the observation that glasses of higher density (formed through pressurization) and glasses of higher stability (formed through slower cooling) also have suppressed boson peak intensities[61-64].

Based on the increase of the storage component during annealing (Figure 2), we can estimate the extent of the boson peak's recovery for the most stable MMT glasses. We estimate the contribution of the boson peak to the dielectric susceptibility based on very high frequency dielectric measurements performed on another small organic molecular glass-former, glycerol[59]; this estimate is reflected in the green "rejuvenated" spectra in Figure 6. Next, we account for the expected suppression of the boson peak in the stable glass. Based on the decrease in the height of the boson peak observed for stable PVD glasses of indomethacin[49], we estimate that the height of the boson peak for a stable glass of MMT is 34% less than the height of the rejuvenated sample. This suppressed boson peak is depicted by the blue "stable" spectra in Figure 6. This is a reasonable extent of suppression, as lowering the cooling rate for organic glasses has been previously observed to result in a boson peak suppression of 7%[63]; we expect a much larger effect for PVD glasses due to their much lower fictive temperatures.

Using these values, we can recreate the 2% increase in the magnitude of $\chi'$ observed at 20 Hz for the longest annealing time for the most stable glass (seen in Figure 2) as shown by the red-dashed "annealed" spectra in Figure 6. The annealed glass spectra has a boson peak intensity that has increased by 7% relative to the stable glass value. Stated another way, when the boson peak recovers 20% of the way to its value in a fresh liquid-cooled glass, the value of $\chi'$ also recovers 20% of the way to its rejuvenated value. According to this scenario, in our frequency range, such a change in $\chi'$ would occur with negligible change in $\chi''$, consistent with our experimental result.

We expect that this explanation for the increase in $\chi'$ without an increase in $\chi''$ (over the frequency window from $1 - 10^5$ Hz) is basically correct. We note that *any process* (not necessarily the boson peak) occurring at a frequency greater than $10^5$ Hz would be a possible candidate for the behavior of $\chi'$ and $\chi''$. However, this explanation leads to a new question: Why would the boson peak (or another high frequency process) recover substantially in ~$10^4$ s at $0.98T_g$ when full rejuvenation requires a much longer time, as indicated by the negligible change in $T_{ons}$ and the estimated rejuvenation time of ~$10^6$ s (Supplemental Figure 8)?

Our observations have some elements in common with a previously reported 'softening' mechanism in the transformation of stable PVD glasses annealed above $T_g$[41, 65], as reported by Vila-Costa et al. These authors performed fast-scanning nanocalorimetry experiments on stable glasses of an organic semiconductor. Thin films of the stable glass were capped by a glass with higher $T_g$ in order to eliminate transformation originating at the free surface. The stable glasses were annealed near $T_g$ + 30 K for various times, after which the transformation was assessed in a temperature ramp (qualitatively as shown in Figure 1). They observed two glass transitions in the annealed films, with the lower temperature transition associated with the supercooled liquid produced by annealing, and a higher temperature transition associated with the stable glass. The higher temperature transition moved to slightly lower temperatures with annealing, which the authors attribute to a partial rejuvenation or softening of the stable glass. Ref [41] reports that the properties of the stable glass can evolve prior to its transformation in the liquid, and at a qualitative level this matches our observations. Quantitatively, there is an important difference, as we observe annealing-induced changes in the dielectric storage at much earlier stages of the transformation, before there is any shift in the onset temperature. Thus, we conclude the apparent softening of the stable glass reported here is distinct from that reported in Ref. [41].

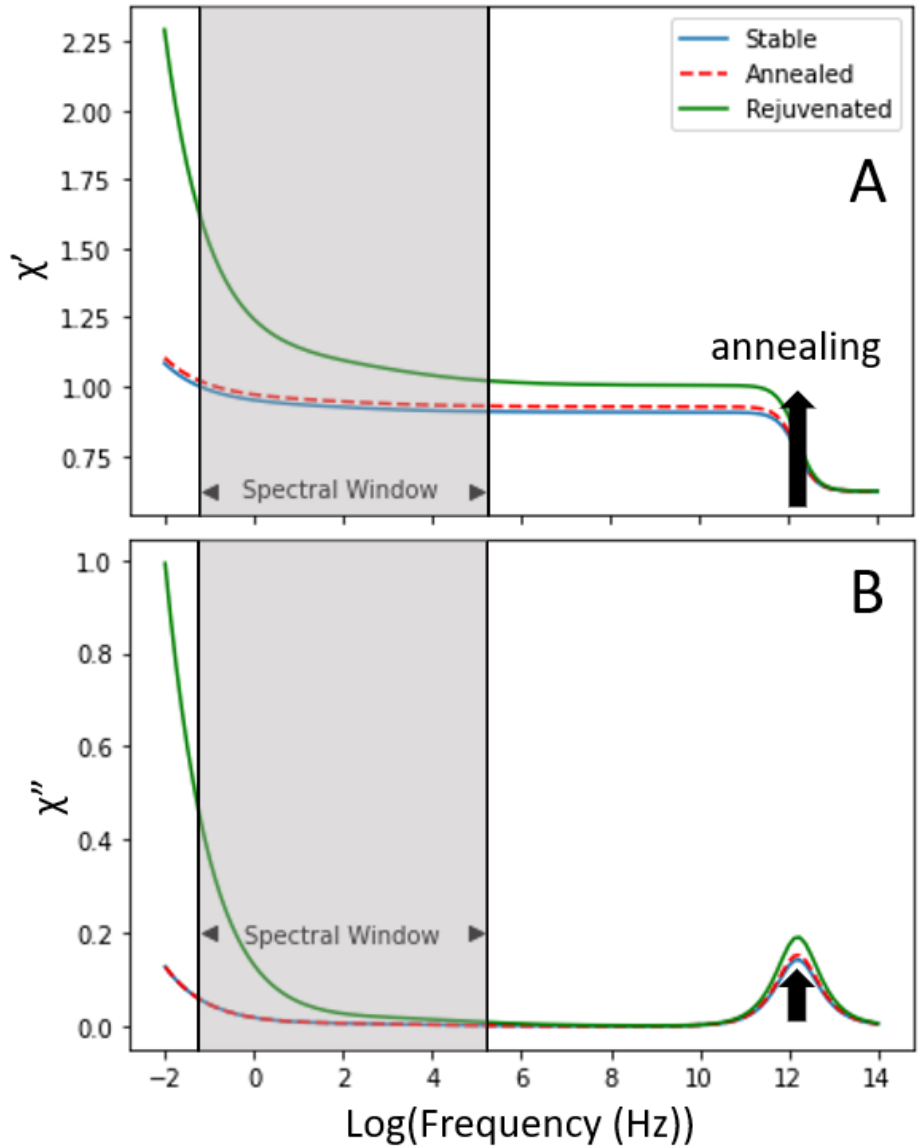

Figure 6: Simulated dielectric response of stable, annealed, and rejuvenated glasses of MMT. The blue spectra represent the stable glass with a suppressed boson peak. In this scenario, a partial recovery of the boson peak leads to an increase in $\chi'$ without an increase in $\chi''$, as observed in Figure 2.

### B. Comparison to Simulations

The experimental results have surprising aspects and we turn to computer simulations for further insights. Simulations based upon the swap Monte Carlo algorithm have reproduced many of the qualitative features of highly stable PVD glasses, including low energy/enthalpy[23], high density[39, 74], transformation fronts at free surfaces[75], and a bulk "nucleation and growth" transformation process[39]. Some features of the simulations, such as the velocity of the transformation front, are in quantitative agreement with experiment.

For this comparison, we make use of a two-dimensional system of polydisperse disks that was equilibrated using the swap Monte Carlo algorithm, at a temperature $T_i = 0.035 \sim T_g/2$. Constant pressure molecular dynamics simulations are performed starting with these configurations. In previous work, it was shown that when such configurations are quickly heated to an annealing temperature near $T_a = 0.1 \sim 1.4 T_g$, transformation occurs via a nucleation and growth process in the bulk[39]. Figure 7 sets the stage for our comparison with the experiments. The upper panel shows how the density of the system evolves at the annealing temperature. The initial drop (at $t = 10^1$) is due to thermal expansion of the glassy solid. The second drop (at $t = 10^7$) is due to transformation to the (less dense) supercooled liquid. The panels at the bottom of Figure 7 show snapshots of configurations at the very earliest stages of the transformation. Mobile particles are those that have exchanged nearest neighbors and are shown in red. Immobile particles (the stable glass) have retained their nearest neighbors and are shown in blue. The fraction of the system that is liquid is indicated for each snapshot.

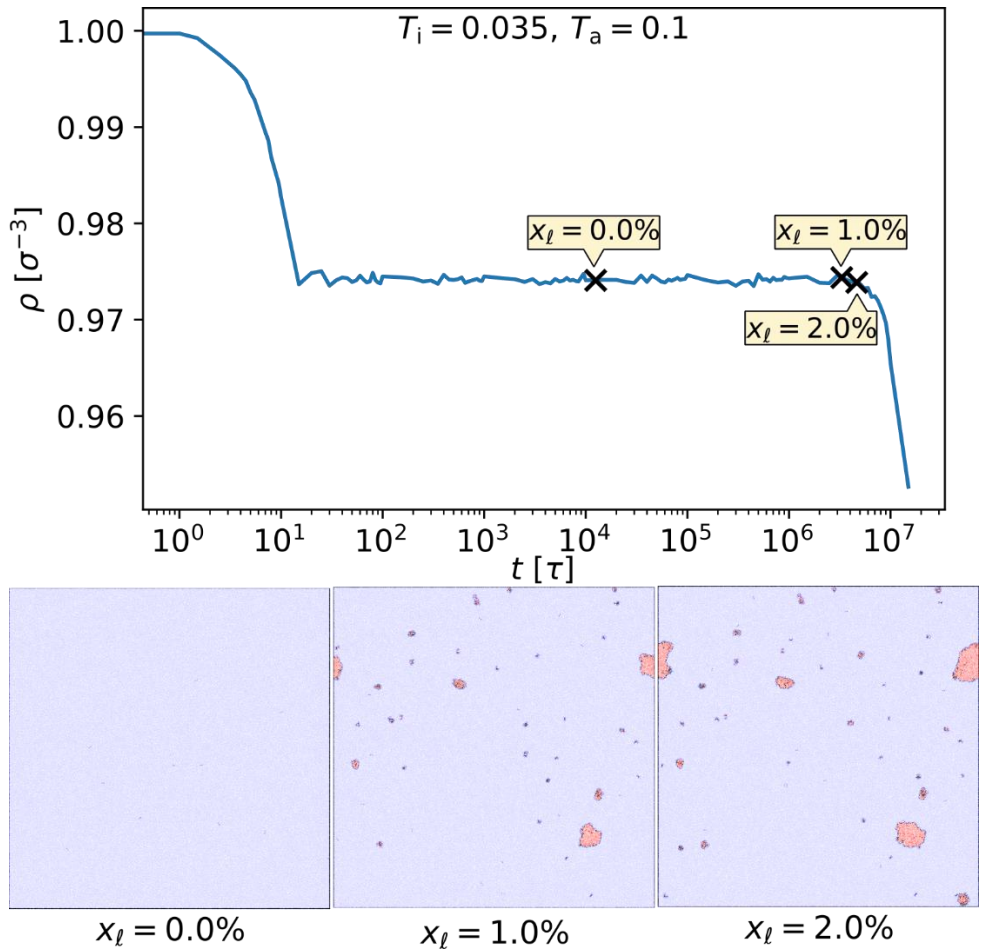

Figure 7: Density as a function of time during the isothermal annealing of a stable computer glass at constant pressure (upper panel). Lower panels: Configurations showing immobile particles (stable glass, in blue) and mobile particles (liquid, in red) at three times after the start of annealing, with $x_l$ indicating the fraction of liquid in the system.

Because the simulations provide a high resolution view of the rejuvenation process, we can directly measure the fraction $x_l(t_w)$ of particles that have already transformed into the supercooled liquid state, by defining mobile particles as those that have exchanged nearest neighbors since the initial condition at $t_w = 0$. This fraction is indicated in each snapshot in Figure 7, and we have studied systems for which $x_l(t_w)$ is smaller than 5%.

For comparison with the experiments, the central issue is whether the properties of the stable glass (blue regions in Figure 7) soften or change in any way during annealing. To this end, we analyze the dynamics of the system starting from selected initial conditions with $x_l =$ 0, 0.5, 1.0, 2.0, and 5%. We have measured the mean-squared displacement (MSD), as defined in Eq. (1), for these five waiting times which correspond to $t_w/(10^6) =$ 0, 3.06, 3.67, 4.59 and 6.12. The measured MSD can be decomposed between liquid and glass particles, corresponding to red and blue particles at time $t_w$ in the snapshots of Fig. 7. In Fig. 8, for each $t_w$, we represent the evolution of the MSD for the glass particles as a function of t. In this plot, we access the single particle dynamics of the glass particles over a broad range of timescales encompassing the vibrational dynamics at short times and the rejuvenation of the $\alpha$ relaxation at long times.

It is evident from Figure 8 that the MSD curves for the glass particles are identical out to times of at least $10^4 - 10^5$, at which point rejuvenation of the $\alpha$ process starts and some initially immobile particles start to transform. Crucially, the short-time vibrational dynamics of the glass (including the Boson peak), which can be seen in the time window from t = 0.1 to t = 100 in Figure 8, is not affected by the emerging liquid regions. Any modification of the glass would affect the peak near t ~ 0.5 or the height of the plateau reached after t ~ 10.

Further dynamic tests of the simulation results confirm the conclusion that the global dynamics of the system is consistent with a linear superposition of $x_l$ liquid particles with (1 - $x_l$) glass particles, with glass dynamics being insensitive to $x_l$ over a broad time window. As an example, in Supplemental Figure 10, we show the van Hove function for the glass particles for various values of $t_w$. All these simulation tests indicate that the glass regions do not appreciably soften (or change their properties in any manner) prior to the transformation into the liquid. Thus, based upon the simulations, we reach a conclusion that is apparently inconsistent with the experimental results.

We do not know why the experiments and simulations indicate different conclusions about the softening of the stable glass prior to its transformation, but we will mention some relevant factors. The experiments and simulations utilize similarly stable configurations. The simulations utilized disk-like particles in two dimensions while the experiments studied molecules in three dimensions. We imagine that the difference between molecules and particles may be important. Specifically, most real molecules are capable of internal motions, including major conformational changes, that can be important for relaxation dynamics[48, 76-78]. In the future, swap Monte Carlo simulations may also be possible with molecular systems, and this would allow a check on this speculation. As a further issue, the limitations of computing power required that a much larger temperature upjump was used in the simulations (roughly 3-fold) in comparison to the experiments (roughly 10%). This can be appreciated in Figure 8, where the simulations show about 5 decades in time between initial cage exploration and stable glass transformation. In the experiments, there is roughly 16 decades in time between these events. This allows the possibility of additional relaxation processes impacting the experiments, possibly explaining the different perspectives given by the experiments and simulations presented here. It will be important for the idea of stable glass softening to be tested with other systems and experimental techniques.

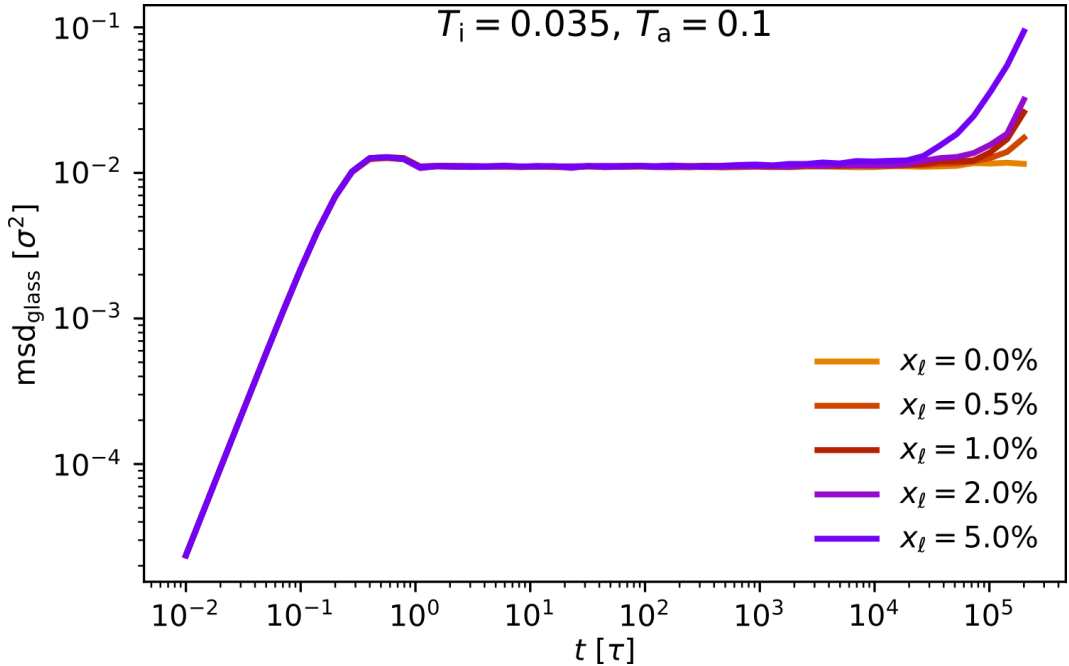

Figure 8: Mean-squared displacement of stable glass particles as a function time, at various times since the beginning of annealing. The dynamic properties of the stable glass do not evolve with annealing in the simulation.

C. Estimation of characteristic relaxation times of stable deposited glasses

At their deposition temperature, highly stable glasses have such long relaxation times that they cannot be directly measured. Estimates of these relaxation times vary from $10^6$ to $10^{27}$ s[79-83], depending upon the system and the method used for the estimate. The measurements shown in Figure 5 provide a new method to estimate these long relaxation times. By making use of the slow initial response of the films to the temperature change, we can extract information about the relaxation times characteristic of the deposited films at the temperature of deposition. We begin this analysis by estimating the initial relaxation times at $0.98T_g$. To accomplish this, we fit the changes in onset temperature shown in Figure 5 to a KWW function with fixed exponent $\beta = 0.5$. The $\tau_{KWW}$ values increase with the stability of the film, from $10^3$ s for the $0.95T_g$ films, to $10^5$ s for the $0.90T_g$ films, to $2 \times 10^7$ s for the $0.85T_g$ films, to $2 \times 10^{10}$ s for the $0.80T_g$ films. These values are plotted as solid symbols in Figure 9. We interpret this trend as the expected increase of the structural relaxation time with decreasing fictive temperature.

We then employ a Tool-Narayanaswamy-Moynihan (TNM) model analysis[20, 84, 85] to extrapolate the KWW relaxation times down to the deposition temperature, making use of material-specific parameters available from the literature. We use the equation:

$$\ln\left(\frac{\tau_{eff}}{\tau_0}\right) = \frac{x * \Delta H_{eff}}{R * T_{meas}} + \frac{(1-x) * \Delta H_{eff}}{R * T_{fictive}} \quad (1)$$

where $\Delta H_{eff}$ is the effective activation enthalpy based on the local slope of the VFT fit for MMT[72], x is the partitioning between the measurement temperature and the fictive temperature which has been found in the past to be equal to 0.32 for MMT[72], and $\tau_0$ is fitted to ensure agreement with the $\tau_{KWW}$ values. Since our measurements do not directly access fictive temperature, the fictive temperature for each glass is treated as a constant, equal to the deposition temperature. Based on previous work[37, 73], this is expected to be a good approximation for the two less stable glasses, but it is likely that the fictive temperature is slightly higher for the two most stable glasses. For the most stable glasses (lower deposition temperatures of $0.85T_g$ and $0.80T_g$), the mobility at the free surface during deposition is enough to enhance the stability of the deposited film, but not sufficient to fully equilibrate the film.

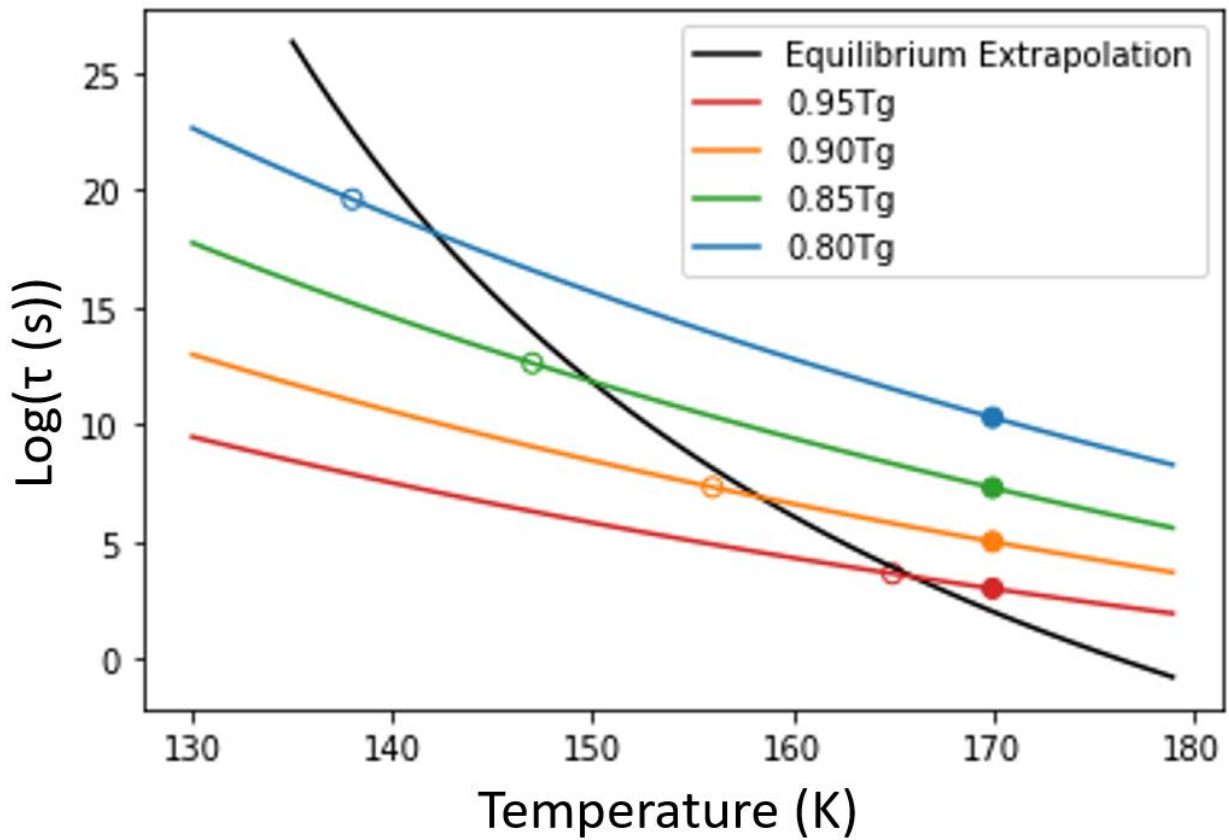

Figure 9: TNM analysis of $\tau_{eff}$ values for glasses of different stabilities (KWW fits from Figure 5, shown here as the solid circles at 170 K) to estimate the relaxation time of each stable glass at the deposition temperature, shown by the hollow circles at the deposition temperature for each iso-fictive temperature line of relaxation times. Extrapolated equilibrium relaxation times are based on a VFT function from the literature[72].

In Figure 9, we show the $\tau_{eff}$ values as a function of temperature, calculated for each glass using Equation 1; we emphasize that these are "iso-fictive temperature" lines. The open symbols indicate the estimated value of $\tau_{eff}$ at the deposition temperature. As seen in Figure 9, the predicted $\tau_{eff}$ value for the $0.95T_g$ glass is very close to that expected (by extrapolation) for the equilibrium supercooled liquid at that deposition temperature. Similarly, the $\tau_{eff}$ value for the $0.90T_g$ glass is close to the value expected for the supercooled liquid. These results are consistent with the view that the VFT temperature dependence of $\tau_\alpha$ above $T_g$ provides a reasonable prediction of $\tau_\alpha$ down to $0.90T_g$.

For the lower deposition temperatures, there is an increasing deviation of the calculated $\tau_{eff}$ value from the VFT-predicted equilibrium relaxation times. This could be interpreted in two ways. First, that the equilibrium relaxation times do not follow the VFT temperature dependence at these lower temperatures. Second, that the films prepared at lower deposition temperatures do not have the same relaxation times as the equilibrium supercooled liquid would at those same temperatures. Since we're fairly confident that the fictive temperatures of these more stable films are not actually equal to the deposition temperature, we lean towards the second interpretation. This does not however rule out the possibility that the equilibrium relaxation times below $0.90T_g$ follow a different temperature dependence than the VFT functional form observed at higher temperatures.

V. Conclusion

In conclusion, by annealing vapor-deposited glasses of MMT slightly below $T_g$, we are able to observe at least the initial stages of rejuvenation into the supercooled liquid. From changes in the onset temperature during annealing, we draw inferences about the relaxation times of glasses of different stabilities. These relaxation times are extremely long for the most stable glasses, as expected from previous experiments where annealing was performed above $T_g$. From changes in the storage component of the dielectric susceptibility, we provide insight about the recovery of the boson peak during annealing. We infer from these experiments that the boson peak recovers more quickly than the transformation of stable glass into supercooled liquid. This would indicate that the properties of the stable glass are evolving prior to its transformation.

We also performed computer simulations to mimic the experimental annealing experiments, starting with ultrastable configurations generated by the swap Monte Carlo method. In the computer simulations, the properties of the stable glass do not evolve prior to transformation, in contradiction with the experimental work. This is surprising given the close correspondence between experimental PVD glasses and swap glasses that has been documented in previous work. Clearly additional experiments and simulations are needed to resolve the question of whether stable glass properties evolve during annealing. One possible approach would be to more directly observe the evolution of the boson peak with neutron or Brillouin light scattering experiments. It would also be useful to perform experiments similar to those reported here on additional molecular systems. Simulations on stable molecular glasses would be particularly useful.

VI. Author Declaration Statements

**Conflicts of Interest Disclosure**

The authors have no conflicts to disclose.

**Author Contributions Statement**

**M. E. Tracy**: Investigation (lead); Conceptualization (equal); Formal Analysis (lead); Visualization (lead); Writing-original draft (lead); Writing-review and editing (equal). **B. J. Kasting**: Conceptualization (equal); Writing-review and editing (supporting). **C. Herrero**: Software (lead); Formal Analysis (lead); Visualization (lead); Writing-original draft (equal). **L. Berthier**: Conceptualization (equal); Writing-review and editing (equal). **R. Richert**: Conceptualization (equal); Writing-review and editing (equal); Funding Acquisition (equal). **A. Guiseppe-Elie**: Resources (supporting); Writing – review and editing (supporting). **M. D. Ediger**: Conceptualization (equal); Funding Acquisition (equal); Writing-review and editing (equal).

VII. Acknowledgements

This work has been supported by the National Science Foundation (CHE-2153944).

References

1. S.-J. Zou, Y. Shen, F.-M. Xie, J.-D. Chen, Y.-Q. Li and J.-X. Tang, *Recent advances in organic light-emitting diodes: toward smart lighting and displays* Materials Chemistry Frontiers **4**, 788-820 (2020). DOI: 10.1039/c9qm00716d.
2. P. G. Debenedetti and F. H. Stillinger, *Supercooled liquids and the glass transition* Nature **410**, 259-67 (2001). DOI: 10.1038/35065704.
3. K. Bagchi, M. E. Fiori, C. Bishop, M. F. Toney and M. D. Ediger, *Stable Glasses of Organic Semiconductor Resist Crystallization* J Phys Chem B **125**, 461-466 (2021). DOI: 10.1021/acs.jpcb.0c09925.
4. G. J. Hedley, A. Ruseckas and I. D. Samuel, *Light Harvesting for Organic Photovoltaics* Chem Rev **117**, 796-837 (2017). DOI: 10.1021/acs.chemrev.6b00215.
5. M. Ashby and A. Greer, *Metallic glasses as structural materials* Scripta Materialia **54**, 321-326 (2006). DOI: 10.1016/j.scriptamat.2005.09.051.
6. W. H. Wang, *Bulk Metallic Glasses with Functional Physical Properties* Advanced Materials **21**, 4524-4544 (2009). DOI: 10.1002/adma.200901053.
7. P. Luo, C. R. Cao, F. Zhu, Y. M. Lv, Y. H. Liu, P. Wen, H. Y. Bai, G. Vaughan, M. di Michiel, B. Ruta and W. H. Wang, *Ultrastable metallic glasses formed on cold substrates* Nat Commun **9**, 1389 (2018). DOI: 10.1038/s41467-018-03656-4.
8. J. Q. Wang, Y. Shen, J. H. Perepezko and M. D. Ediger, *Increasing the kinetic stability of bulk metallic glasses* Acta Materialia **104**, 25-32 (2016). DOI: 10.1016/j.actamat.2015.11.048.
9. H. B. Yu, Y. Luo and K. Samwer, *Ultrastable metallic glass* Adv Mater **25**, 5904-8 (2013). DOI: 10.1002/adma.201302700.
10. G. Biroli and J. P. Garrahan, *Perspective: The glass transition* J Chem Phys **138**, 12A301 (2013). DOI: 10.1063/1.4795539.
11. R. Casalini and C. M. Roland, *Aging of the secondary relaxation to probe structural relaxation in the glassy state* Phys Rev Lett **102**, 035701 (2009). DOI: 10.1103/PhysRevLett.102.035701.
12. E. Duval, L. Saviot, L. David, S. Etienne and J. F. Jal, *Effect of physical aging on the low-frequency vibrational density of states of a glassy polymer* Europhysics Letters (EPL) **63**, 778-784 (2003). DOI: 10.1209/epl/i2003-00573-x.
13. T. D. Jaeger and D. S. Simmons, *Temperature dependence of aging dynamics in highly non-equilibrium model polymer glasses* J Chem Phys **156**, 114504 (2022). DOI: 10.1063/5.0080717.
14. V. Lubchenko and P. G. Wolynes, *Theory of aging in structural glasses* J Chem Phys **121**, 2852-65 (2004). DOI: 10.1063/1.1771633.
15. X. Monnier, S. Marina, X. Lopez de Pariza, H. Sardon, J. Martin and D. Cangialosi, *Physical Aging Behavior of a Glassy Polyether* Polymers (Basel) **13** (2021). DOI: 10.3390/polym13060954.
16. B. Riechers, L. A. Roed, S. Mehri, T. S. Ingebrigtsen, T. Hecksher, J. C. Dyre and K. Niss, *Predicting nonlinear physical aging of glasses from equilibrium relaxation via the material time* Sci Adv **8**, eabl9809 (2022). DOI: 10.1126/sciadv.abl9809.
17. Y. Zhao, B. Shang, B. Zhang, X. Tong, H. Ke, H. Bai and W. H. Wang, *Ultrastable metallic glass by room temperature aging* Sci Adv **8**, eabn3623 (2022). DOI: 10.1126/sciadv.abn3623.
18. M. D. Ediger, C. A. Angell and S. R. Nagel, *Supercooled Liquids and Glasses* The Journal of Physical Chemistry **100**, 13200-13212 (1996). DOI: 10.1021/jp953538d.

19. R. Bruning and K. Samwer, *Glass transition on long time scales* Phys Rev B Condens Matter **46**, 11318-11322 (1992). DOI: 10.1103/physrevb.46.11318.
20. C. T. Moynihan, A. J. Easteal, M. A. Bolt and J. Tucker, *Dependence of the Fictive Temperature of Glass on Cooling Rate* Journal of the American Ceramic Society **59**, 12-16 (1976). DOI: 10.1111/j.1151-2916.1976.tb09376.x.
21. K. Kawakami, *Ultraslow Cooling for the Stabilization of Pharmaceutical Glasses* J Phys Chem B **123**, 4996-5003 (2019). DOI: 10.1021/acs.jpcb.9b02122.
22. S. F. Swallen, K. L. Kearns, M. K. Mapes, Y. S. Kim, R. J. McMahon, M. D. Ediger, T. Wu, L. Yu and S. Satija, *Organic glasses with exceptional thermodynamic and kinetic stability* Science **315**, 353-6 (2007). DOI: 10.1126/science.1135795.
23. L. Berthier, P. Charbonneau, E. Flenner and F. Zamponi, *Origin of Ultrastability in Vapor-Deposited Glasses* Phys Rev Lett **119**, 188002 (2017). DOI: 10.1103/PhysRevLett.119.188002.
24. Y. Chen, M. Zhu, A. Laventure, O. Lebel, M. D. Ediger and L. Yu, *Influence of Hydrogen Bonding on the Surface Diffusion of Molecular Glasses: Comparison of Three Triazines* J Phys Chem B **121**, 7221-7227 (2017). DOI: 10.1021/acs.jpcb.7b05333.
25. A. Laventure, A. Gujral, O. Lebel, C. Pellerin and M. D. Ediger, *Influence of Hydrogen Bonding on the Kinetic Stability of Vapor-Deposited Glasses of Triazine Derivatives* J Phys Chem B **121**, 2350-2358 (2017). DOI: 10.1021/acs.jpcb.6b12676.
26. P. Luo and Z. Fakhraai, *Surface-Mediated Formation of Stable Glasses* Annu Rev Phys Chem **74**, 361-389 (2023). DOI: 10.1146/annurev-physchem-042018-052708.
27. M. Tylinski, M. S. Beasley, Y. Z. Chua, C. Schick and M. D. Ediger, *Limited surface mobility inhibits stable glass formation for 2-ethyl-1-hexanol* J Chem Phys **146**, 203317 (2017). DOI: 10.1063/1.4977787.
28. Y. Zhang and Z. Fakhraai, *Decoupling of surface diffusion and relaxation dynamics of molecular glasses* Proc Natl Acad Sci U S A **114**, 4915-4919 (2017). DOI: 10.1073/pnas.1701400114.
29. S. Samanta, G. Huang, G. Gao, Y. Zhang, A. Zhang, S. Wolf, C. N. Woods, Y. Jin, P. J. Walsh and Z. Fakhraai, *Exploring the Importance of Surface Diffusion in Stability of Vapor-Deposited Organic Glasses* J Phys Chem B **123**, 4108-4117 (2019). DOI: 10.1021/acs.jpcb.9b01012.
30. E. Thoms, J. P. Gabriel, A. Guiseppi-Elie, M. D. Ediger and R. Richert, *In situ observation of fast surface dynamics during the vapor-deposition of a stable organic glass* Soft Matter **16**, 10860-10864 (2020). DOI: 10.1039/d0sm01916j.
31. S. S. Dalal, Z. Fakhraai and M. D. Ediger, *High-throughput ellipsometric characterization of vapor-deposited indomethacin glasses* J Phys Chem B **117**, 15415-25 (2013). DOI: 10.1021/jp405005n.
32. S. S. Dalal, D. M. Walters, I. Lyubimov, J. J. de Pablo and M. D. Ediger, *Tunable molecular orientation and elevated thermal stability of vapor-deposited organic semiconductors* Proc Natl Acad Sci U S A **112**, 4227-32 (2015). DOI: 10.1073/pnas.1421042112.
33. M. D. Ediger, *Perspective: Highly stable vapor-deposited glasses* J Chem Phys **147**, 210901 (2017). DOI: 10.1063/1.5006265.
34. B. J. Kasting, M. S. Beasley, A. Guiseppi-Elie, R. Richert and M. D. Ediger, *Relationship between aged and vapor-deposited organic glasses: Secondary relaxations in methyl-m-toluate* J Chem Phys **151**, 144502 (2019). DOI: 10.1063/1.5123305.

35. K. J. Dawson, L. Zhu, L. Yu and M. D. Ediger, *Anisotropic structure and transformation kinetics of vapor-deposited indomethacin glasses* J Phys Chem B **115**, 455-63 (2011). DOI: 10.1021/jp1092916.
36. S. F. Swallen, K. Traynor, R. J. McMahon, M. D. Ediger and T. E. Mates, *Stable glass transformation to supercooled liquid via surface-initiated growth front* Phys Rev Lett **102**, 065503 (2009). DOI: 10.1103/PhysRevLett.102.065503.
37. M. Tylinski, A. Sepulveda, D. M. Walters, Y. Z. Chua, C. Schick and M. D. Ediger, *Vapor-deposited glasses of methyl-m-toluate: How uniform is stable glass transformation?* J Chem Phys **143**, 244509 (2015). DOI: 10.1063/1.4938420.
38. A. Sepulveda, S. F. Swallen and M. D. Ediger, *Manipulating the properties of stable organic glasses using kinetic facilitation* J Chem Phys **138**, 12A517 (2013). DOI: 10.1063/1.4772594.
39. C. Herrero, C. Scalliet, M. D. Ediger and L. Berthier, *Two-step devitrification of ultrastable glasses* Proc Natl Acad Sci U S A **120**, e2220824120 (2023). DOI: 10.1073/pnas.2220824120.
40. M. Ruiz-Ruiz, A. Vila-Costa, T. Bar, C. Rodríguez-Tinoco, M. Gonzalez-Silveira, J. A. Plaza, J. Alcalá, J. Fraxedas and J. Rodriguez-Viejo, *Real-time microscopy of the relaxation of a glass* Nature Physics (2023). DOI: 10.1038/s41567-023-02125-0.
41. A. Vila-Costa, M. Gonzalez-Silveira, C. Rodríguez-Tinoco, M. Rodríguez-López and J. Rodriguez-Viejo, *Emergence of equilibrated liquid regions within the glass* Nature Physics **19**, 114-119 (2022). DOI: 10.1038/s41567-022-01791-w.
42. B. J. Kasting, J. P. Gabriel, M. E. Tracy, A. Guiseppi-Elie, R. Richert and M. D. Ediger, *Unusual Transformation of Mixed Isomer Decahydroisoquinoline Stable Glasses* J Phys Chem B (2023). DOI: 10.1021/acs.jpcb.3c00459.
43. C. Herrero, M. D. Ediger and L. Berthier, *Front propagation in ultrastable glasses is dynamically heterogeneous* J Chem Phys **159** (2023). DOI: 10.1063/5.0168506.
44. A. Ninarello, L. Berthier and D. Coslovich, *Models and Algorithms for the Next Generation of Glass Transition Studies* Physical Review X **7** (2017). DOI: 10.1103/PhysRevX.7.021039.
45. C. Scalliet, B. Guiselin and L. Berthier, *Thirty Milliseconds in the Life of a Supercooled Liquid* Physical Review X **12** (2022). DOI: 10.1103/PhysRevX.12.041028.
46. L. Berthier, P. Charbonneau, D. Coslovich, A. Ninarello, M. Ozawa and S. Yaida, *Configurational entropy measurements in extremely supercooled liquids that break the glass ceiling* Proc Natl Acad Sci U S A **114**, 11356-11361 (2017). DOI: 10.1073/pnas.1706860114.
47. H. B. Yu, M. Tylinski, A. Guiseppi-Elie, M. D. Ediger and R. Richert, *Suppression of beta Relaxation in Vapor-Deposited Ultrastable Glasses* Phys Rev Lett **115**, 185501 (2015). DOI: 10.1103/PhysRevLett.115.185501.
48. C. Rodriguez-Tinoco, K. L. Ngai, M. Rams-Baron, J. Rodriguez-Viejo and M. Paluch, *Distinguishing different classes of secondary relaxations from vapour deposited ultrastable glasses* Phys Chem Chem Phys **20**, 21925-21933 (2018). DOI: 10.1039/c8cp02341g.
49. T. Perez-Castaneda, C. Rodriguez-Tinoco, J. Rodriguez-Viejo and M. A. Ramos, *Suppression of tunneling two-level systems in ultrastable glasses of indomethacin* Proc Natl Acad Sci U S A **111**, 11275-80 (2014). DOI: 10.1073/pnas.1405545111.
50. D. Khomenko, C. Scalliet, L. Berthier, D. R. Reichman and F. Zamponi, *Depletion of Two-Level Systems in Ultrastable Computer-Generated Glasses* Phys Rev Lett **124**, 225901 (2020). DOI: 10.1103/PhysRevLett.124.225901.

51. M. Gonzalez-Jimenez, T. Barnard, B. A. Russell, N. V. Tukachev, U. Javornik, L. A. Hayes, A. J. Farrell, S. Guinane, H. M. Senn, A. J. Smith, M. Wilding, G. Mali, M. Nakano, Y. Miyazaki, P. McMillan, G. C. Sosso and K. Wynne, *Understanding the emergence of the boson peak in molecular glasses* Nat Commun **14**, 215 (2023). DOI: 10.1038/s41467-023-35878-6.
52. T. S. Grigera, V. Martin-Mayor, G. Parisi and P. Verrocchio, *Phonon interpretation of the 'boson peak' in supercooled liquids* Nature **422**, 289-92 (2003). DOI: 10.1038/nature01475.
53. Y.-C. Hu and H. Tanaka, *Origin of the boson peak in amorphous solids* Nature Physics **18**, 669-677 (2022). DOI: 10.1038/s41567-022-01628-6.
54. H. Shintani and H. Tanaka, *Universal link between the boson peak and transverse phonons in glass* Nat Mater **7**, 870-7 (2008). DOI: 10.1038/nmat2293.
55. H. P. Zhang, B. B. Fan, J. Q. Wu, W. H. Wang and M. Z. Li, *Universal relationship of boson peak with Debye level and Debye-Waller factor in disordered materials* Physical Review Materials **4** (2020). DOI: 10.1103/PhysRevMaterials.4.095603.
56. T. Perez-Castaneda, R. J. Jimenez-Rioboo and M. A. Ramos, *Two-level systems and boson peak remain stable in 110-million-year-old amber glass* Phys Rev Lett **112**, 165901 (2014). DOI: 10.1103/PhysRevLett.112.165901.
57. T. Pérez-Castañeda, R. J. Jiménez-Riobóo and M. A. Ramos, *Do two-level systems and boson peak persist or vanish in hyperaged geological glasses of amber?* Philosophical Magazine **96**, 774-787 (2015). DOI: 10.1080/14786435.2015.1111530.
58. M. A. Ramos, T. Pérez-Castañeda, R. J. Jiménez-Riobóo, C. Rodríguez-Tinoco and J. Rodríguez-Viejo, *Do tunneling states and boson peak persist or disappear in extremely stabilized glasses?* Low Temperature Physics **41**, 412-418 (2015). DOI: 10.1063/1.4922089.
59. P. Lunkenheimer and A. Loidl, *Dielectric spectroscopy of glass-forming materials: α-relaxation and excess wing* Chemical Physics **284**, 205-219 (2002). DOI: 10.1016/s0301-0104(02)00549-9.
60. B. Ruta, G. Baldi, V. M. Giordano, L. Orsingher, S. Rols, F. Scarponi and G. Monaco, *Communication: High-frequency acoustic excitations and boson peak in glasses: A study of their temperature dependence* J Chem Phys **133**, 041101 (2010). DOI: 10.1063/1.3460815.
61. K. Niss, B. Begen, B. Frick, J. Ollivier, A. Beraud, A. Sokolov, V. N. Novikov and C. Alba-Simionesco, *Influence of pressure on the boson peak: stronger than elastic medium transformation* Phys Rev Lett **99**, 055502 (2007). DOI: 10.1103/PhysRevLett.99.055502.
62. A. Monaco, A. I. Chumakov, G. Monaco, W. A. Crichton, A. Meyer, L. Comez, D. Fioretto, J. Korecki and R. Ruffer, *Effect of densification on the density of vibrational states of glasses* Phys Rev Lett **97**, 135501 (2006). DOI: 10.1103/PhysRevLett.97.135501.
63. R. Calemczuk, R. Lagnier and E. Bonjour, *Specific heat of glycerol: Crystalline and glassy states* Journal of Non-Crystalline Solids **34**, 149-152 (1979). DOI: 10.1016/0022-3093(79)90014-0.
64. Y. Gao, C. Yang, G. Ding, L.-H. Dai and M.-Q. Jiang, *Structural rejuvenation of a well-aged metallic glass* Fundamental Research (2022). DOI: 10.1016/j.fmre.2022.12.004.
65. A. Vila-Costa, J. Rafols-Ribe, M. Gonzalez-Silveira, A. F. Lopeandia, L. Abad-Munoz and J. Rodriguez-Viejo, *Nucleation and Growth of the Supercooled Liquid Phase Control Glass Transition in Bulk Ultrastable Glasses* Phys Rev Lett **124**, 076002 (2020). DOI: 10.1103/PhysRevLett.124.076002.
66. R. Richert, *A simple current-to-voltage interface for dielectric relaxation measurements in the range $10^{-3}$ to $10^7$ Hz* Review of Scientific Instruments **67**, 3217-3221 (1996). DOI: 10.1063/1.1147445.

67. L. P. Singh and R. Richert, *Two-channel impedance spectroscopy for the simultaneous measurement of two samples* Rev Sci Instrum **83**, 033903 (2012). DOI: 10.1063/1.3697732.
68. B. Riechers, A. Guiseppi-Elie, M. D. Ediger and R. Richert, *Ultrastable and polyamorphic states of vapor-deposited 2-methyltetrahydrofuran* J Chem Phys **150**, 214502 (2019). DOI: 10.1063/1.5091796.
69. A. J. Kovacs, Transition vitreuse dans les polymères amorphes. Etude phénoménologique. In *Fortschritte der hochpolymeren-forschung*, Springer: 2006; pp 394-507.
70. S. Cheng, Y. Lee, J. Yu, L. Yu and M. D. Ediger, *Surface Equilibration Mechanism Controls the Stability of a Model Codeposited Glass Mixture of Organic Semiconductors* J Phys Chem Lett **14**, 4297-4303 (2023). DOI: 10.1021/acs.jpclett.3c00728.
71. C. Rodríguez-Tinoco, M. Gonzalez-Silveira, J. Ràfols-Ribé, G. Garcia and J. Rodríguez-Viejo, *Highly stable glasses of celecoxib: Influence on thermo-kinetic properties, microstructure and response towards crystal growth* Journal of Non-Crystalline Solids **407**, 256-261 (2015). DOI: 10.1016/j.jnoncrysol.2014.07.031.
72. Z. Chen, Y. Zhao and L. M. Wang, *Enthalpy and dielectric relaxations in supercooled methyl m-toluate* J Chem Phys **130**, 204515 (2009). DOI: 10.1063/1.3142142.
73. A. Sepulveda, M. Tylinski, A. Guiseppi-Elie, R. Richert and M. D. Ediger, *Role of fragility in the formation of highly stable organic glasses* Phys Rev Lett **113**, 045901 (2014). DOI: 10.1103/PhysRevLett.113.045901.
74. C. J. Fullerton and L. Berthier, *Density controls the kinetic stability of ultrastable glasses* EPL (Europhysics Letters) **119** (2017). DOI: 10.1209/0295-5075/119/36003.
75. E. Flenner, L. Berthier, P. Charbonneau and C. J. Fullerton, *Front-Mediated Melting of Isotropic Ultrastable Glasses* Phys Rev Lett **123**, 175501 (2019). DOI: 10.1103/PhysRevLett.123.175501.
76. K. L. Ngai and M. Paluch, *Classification of secondary relaxation in glass-formers based on dynamic properties* J Chem Phys **120**, 857-73 (2004). DOI: 10.1063/1.1630295.
77. M. Paluch, S. Pawlus, S. Hensel-Bielowka, E. Kaminska, D. Prevosto, S. Capaccioli, P. A. Rolla and K. L. Ngai, *Two secondary modes in decahydroisoquinoline: which one is the true Johari Goldstein process?* J Chem Phys **122**, 234506 (2005). DOI: 10.1063/1.1931669.
78. P. Wlodarczyk, B. Czarnota, M. Paluch, S. Pawlus and J. Ziolo, *Microscopic origin of secondary modes observed in decahydroisoquinoline* Journal of Molecular Structure **975**, 200-204 (2010). DOI: 10.1016/j.molstruc.2010.04.023.
79. L. Berthier and M. D. Ediger, *How to "measure" a structural relaxation time that is too long to be measured?* J Chem Phys **153**, 044501 (2020). DOI: 10.1063/5.0015227.
80. G. B. McKenna, *Diverging views on glass transition* Nature Physics **4**, 673-673 (2008). DOI: 10.1038/nphys1063.
81. G. B. McKenna and J. Zhao, *Accumulating evidence for non-diverging time-scales in glass-forming fluids* Journal of Non-Crystalline Solids **407**, 3-13 (2015). DOI: 10.1016/j.jnoncrysol.2014.08.012.
82. D. Kong, Y. Meng and G. B. McKenna, *Searching for the ideal glass transition: Going to yotta seconds and beyond* Journal of Non-Crystalline Solids **606** (2023). DOI: 10.1016/j.jnoncrysol.2023.122186.
83. K. L. Kearns, S. F. Swallen, M. D. Ediger, T. Wu, Y. Sun and L. Yu, *Hiking down the energy landscape: progress toward the Kauzmann temperature via vapor deposition* J Phys Chem B **112**, 4934-42 (2008). DOI: 10.1021/jp7113384.

84. O. S. Narayanaswamy, *A Model of Structural Relaxation in Glass* Journal of the American Ceramic Society **54**, 491-498 (1971). DOI: 10.1111/j.1151-2916.1971.tb12186.x.
85. A. Q. Tool, *Relation between Inelastic Deformability and Thermal Expansion of Glass in Its Annealing Range\** Journal of the American Ceramic Society **29**, 240-253 (1946). DOI: 10.1111/j.1151-2916.1946.tb11592.x.

# Supplementary Material for Initial Stages of Rejuvenation of Vapor-Deposited Glasses during Isothermal Annealing: Contrast Between Experiment and Simulation

M. E. Tracy*,[1], B. J. Kasting[1], C. Herrero[2,3], L. Berthier[2,4], R. Richert[5], A. Guiseppi-Elie[6], M. D. Ediger[1]

[1] *Department of Chemistry, University of Wisconsin-Madison, 1101 University Ave, Madison, Wisconsin 53706, USA*

[2] *Laboratoire Charles Coulomb (L2C), Université de Montpellier, CNRS, 34095 Montpellier, France*

[3] *Institut Laue-Langevin, 71 Avenue des Martyrs, 38042 Grenoble, France*

[4] *Gulliver, UMR CNRS 7083, ESPCI Paris, PSL Research University, 75005 Paris, France*

[5] *School of Molecular Sciences, Arizona State University, Tempe, Arizona 85287, USA*

[6] *Department of Electrical and Computer Engineering, Texas A&M University, College Station, Texas 77843, USA*

* corresponding author: megantracy88@gmail.com

### A. Thickness and deposition rate determination

To determine the thickness of the deposited film of MMT, we turned to a calibration method used recently in a similar vapor-deposition system[30] that relies on the characteristics of the IDE finger spacing, which determines how far above the surface of the substrate molecules of a film contribute to the measured dielectric susceptibility. The minimum thickness to accomplish this is equal to $\lambda/2 = [2(s+w)]/2 = 20$ μm for the IDE's finger geometry. The saturated value of the real component of the dielectric susceptibility once this minimum thickness has been reached can be used to determine the thickness of a film within the thin film limit. Within this regime (for our IDE, this corresponds to thicknesses up to 500 nm), the measured storage corresponds linearly with the thickness of the deposited film according to the following expression:

$$\chi'_{eff} = \chi'_{sat} \frac{h}{\frac{\lambda}{8}} \qquad (S1)$$

where $\chi'_{eff}$ is the measured storage, h is the thickness of the film, $\lambda$ is the characteristic periodicity length for the IDE geometry, and $\chi'_{sat}$ is the measured saturation value of the storage component.

The storage and loss components measured during deposition of a film over 60 μm thick are shown in Supplemental Figure 1. The value of $\chi'_{sat}$ for MMT at this deposition temperature was extracted.

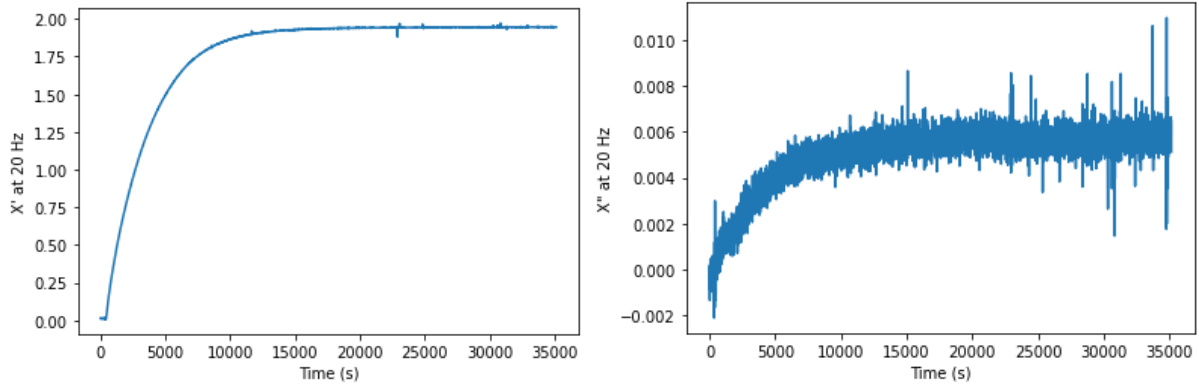

Supplemental Figure 1: Dielectric susceptibility of MMT film during deposition at 0.84Tg for around 9.5 hours.

By fitting the above expression to the data during the first few hundred seconds of this deposition, allowing the deposition rate to vary (assuming a constant deposition rate, such that h = rate * time), this gives a slope of 1.93 nm/s, represented by the orange curve in Supplemental Figure 2.

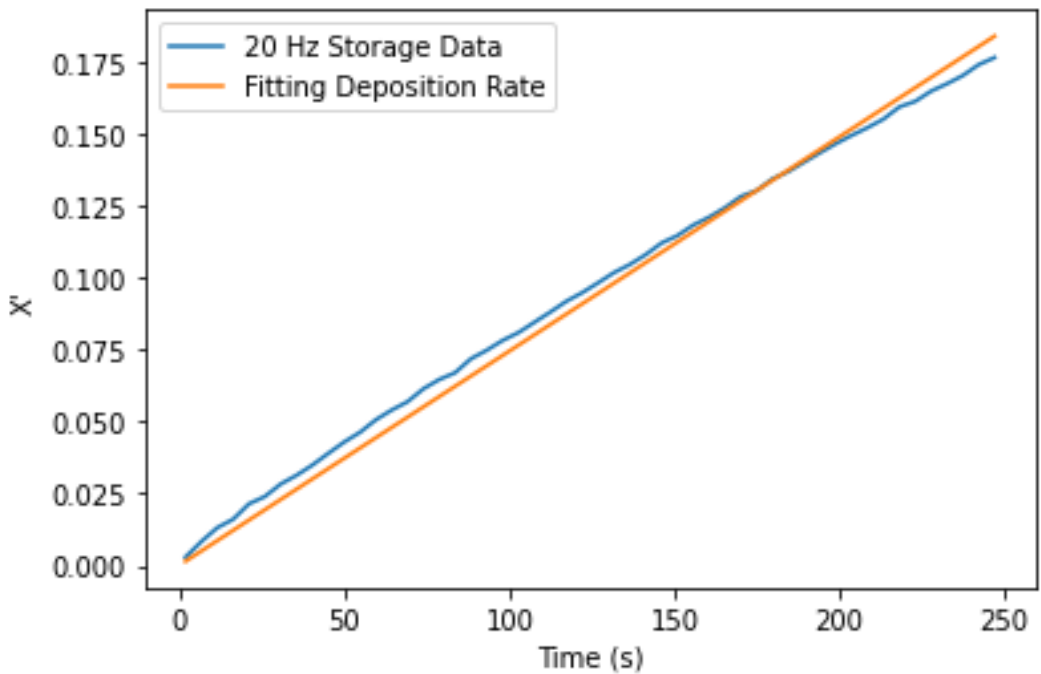

Supplemental Figure 2: Fitting to the first 250 seconds of deposition of the film from Supplemental Figure 1 (experimental data shown in blue) to the expression given above (orange).

The fitted rate gives an estimated total thickness of 66.4 μm, far in excess of the expected thickness needed to fully fill the capacitor. This is in agreement with the observed turnover and long plateau at later times in the experimental data shown in Supplemental Figure 1.

This procedure yields rates that are 3.24 times higher than those estimated by our previous procedure, which relied on a calibration of the change of the voltage differential of a pair of thermopiles on an adjacent finger within the chamber to an ellipsometry measurement of the thickness, as described in a previous publication[37] and utilized in several previous studies. At present, we prefer to rely on the current procedure, as it directly evaluates the film deposited onto the IDE by making use of its fundamental measurement properties.

B. Explanation of onset temperature fitting and distinguishing the bulk transformation process from the surface front transformation mechanism

In this paper, we define the onset temperature somewhat differently than in many previous works. In most cases, the onset temperature marks the initial appearance of the supercooled liquid from a deposited glass on heating. Typically, in uncapped films, this occurs first through a surface-initiated growth front mechanism; however, here we are more interested in the bulk rejuvenation process, which occurs at slightly later times[36, 37, 42, 86, 87]. The bulk rejuvenation process informs us more about the dynamics of the deposited glass, as opposed to the faster surface dynamics.

Empirically, this onset is defined as shown schematically in Supplemental Figure 3. A straight line is fitted to the loss at low temperatures, before the appearance of the surface front in the gradual initial rise beginning at around 180 K. Then, a second line is fitted to the bulk transformation process, which appears as the sharp final approach to the same value as the ordinary supercooled liquid at the same temperature. The temperature where these two lines intersect is defined as the onset temperature in our analysis.

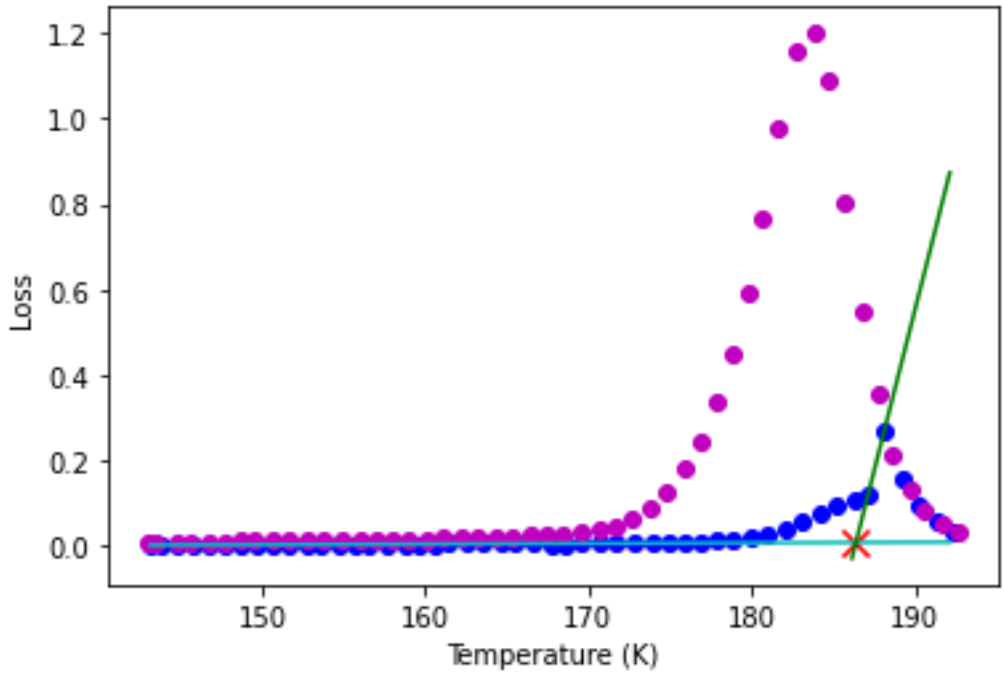

Supplemental Figure 3: Determination of the onset temperature for an annealed film deposited at $0.80T_g$. The teal line is a fit to the low temperature, glassy loss component of dielectric susceptibility, the green line is the fit to the final approach to the supercooled liquid at high temperature, and the red 'x' marks their intersection, defining the onset temperature. The magenta data points are the loss of the ordinary liquid cooled glass on heating.

In this procedure, we ignore the initial rise at lower temperatures (around 180 K in Supplemental Figure 3), which we associate with the surface-initiated growth front transformation mechanism. To be certain of this assignment of the two separate increases in the loss signal during heating to the surface and bulk processes, we performed a series of depositions of most stable $0.80T_g$ films to compare their transformation mechanisms. For the thinnest films, as shown in blue in Supplemental Figure 4, the entire film transforms through the surface mechanism before the bulk process can appreciably contribute. However, with increasing film thickness, the bulk process appears and becomes more and more prominent for the thicker films.

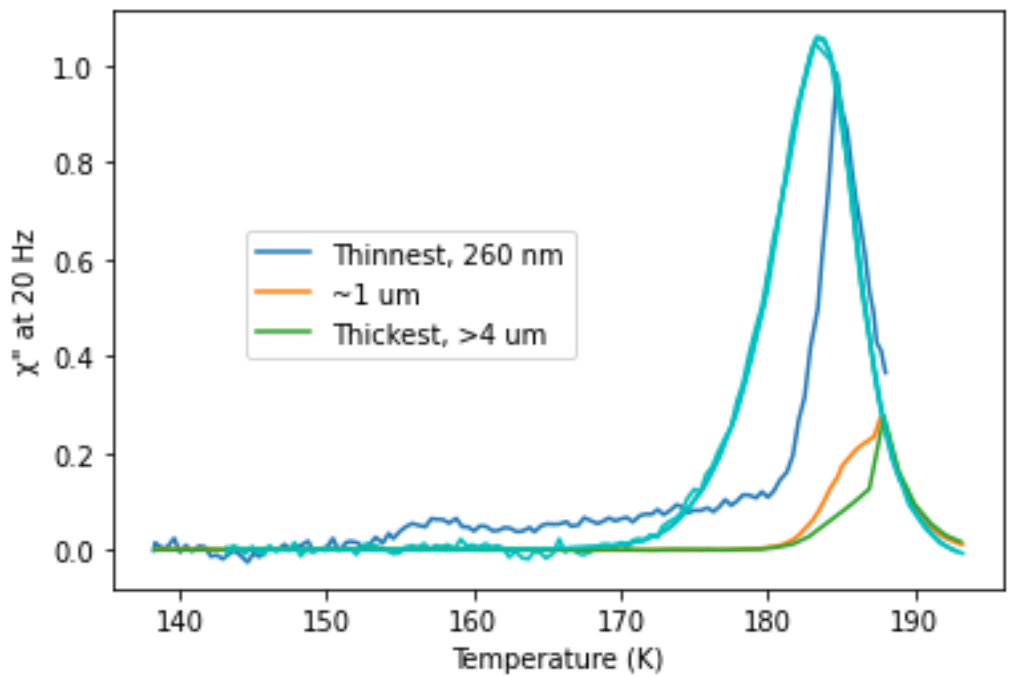

Supplemental Figure 4: Normalized loss during transformation at 5 K/min of most stable $0.80T_g$ glasses of differing thicknesses. The corresponding liquid cooled glass transformations on heating at the same rate are included in cyan.

C. Equivalence of PVD and annealed $0.98T_g$ films and comparison of final state of $0.95T_g$ films annealed at $0.98T_g$ to annealed $0.98T_g$ films

As discussed in the main text, the final steady-state loss value for the somewhat stable $0.95T_g$ film annealed at $0.98T_g$ was 8% lower than the value measured at the annealing temperature during heating of the corresponding liquid cooled glass. We justify this difference based on the fact that the liquid cooled film during heating is not at equilibrium at $0.98T_g$. To confirm this theory, we annealed a liquid cooled glass to equilibrium at $0.98T_g$, and found that the steady state loss value during annealing was 8% lower than the value for the liquid cooled glass on heating. Similarly, the initial value of the loss for films deposited at $0.98T_g$ is 8% lower than the loss of the corresponding liquid cooled glass on heating. The definition of these different loss values is demonstrated schematically in Supplemental Figure 5. Therefore, the final steady-state loss value of the 0.95T$_g$ film annealed at 0.98T$_g$ being 8% below the value of its equivalent liquid cooled glass is in agreement with the glass reaching equilibrium at 0.98T$_g$.

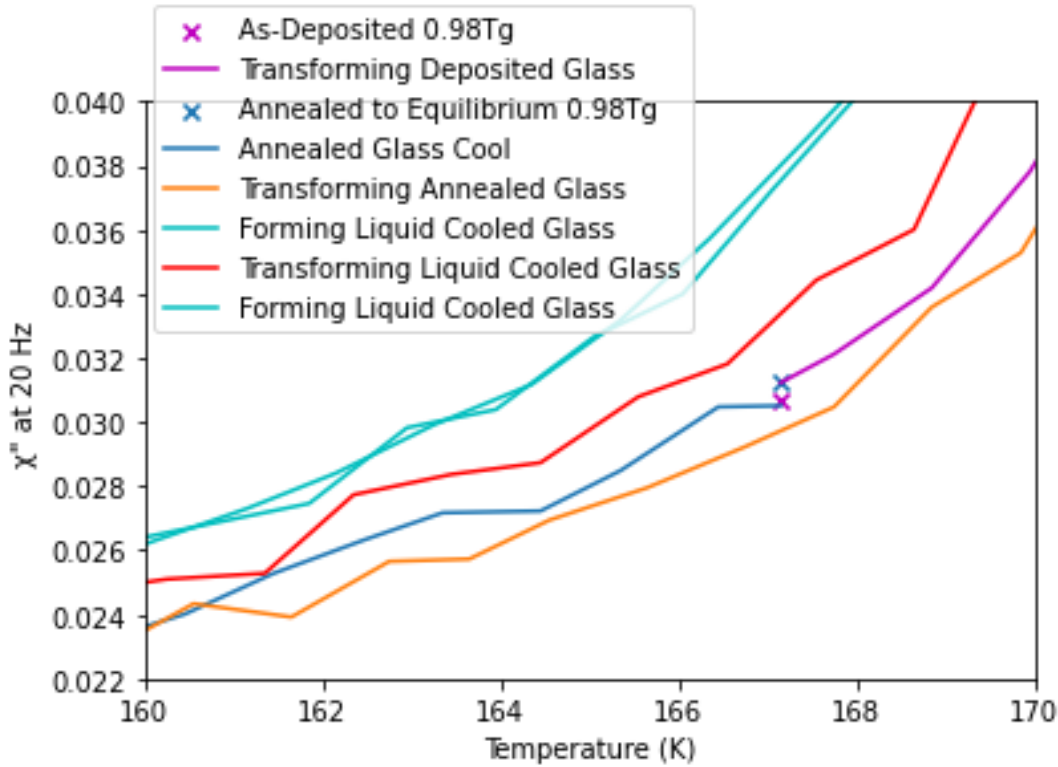

Supplemental Figure 5: Comparison of the initial and annealed loss values for the same film deposited at 0.98T$_g$ (magenta 'x'), transformed at 5 K/min (magenta line), annealed to equilibrium at 0.98T$_g$ (blue 'x'), and then transformed at 5 K/min. The loss value of the as-deposited glass is 8% lower than that of the liquid cooled glass during heating (red line) at the same temperature, while the equilibrium annealed value is 8% lower than the liquid cooled glass value. The as-deposited glass value was defined by the steady state value from the deposition file before the transformation heating began, and the annealed glass value was defined by the steady state value from the annealing file before ramping began. The legend lists the steps in order of time.

Similarly, when we account for differences in thickness, films aged to equilibrium at 0.98T$_g$, regardless of the starting state, reach the same final state at 0.98T$_g$. As seen in Supplemental Figure 6, an ordinary film reaches the same final value for the loss as the PVD film deposited at 0.95T$_g$ after annealing for six hours at 0.98T$_g$. This indicates that the same (or equivalent) state is achieved by annealing to steady state at 0.98T$_g$, regardless of whether the glass starts as a stable deposited glass or not.

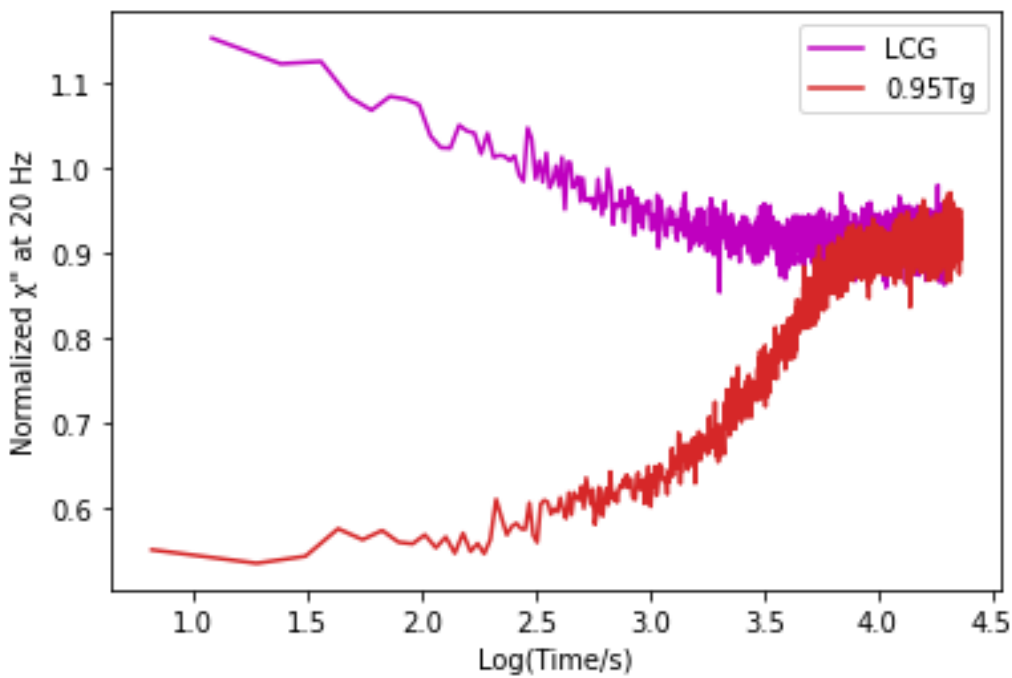

Supplemental Figure 6: Loss during annealing to convergent equilibrium at $0.98T_g$, starting from an ordinary glass cooled at 10 K/min (blue) and from a PVD glass deposited at $0.95T_g$. These curves are normalized to value of the loss at the annealing temperature the corresponding ordinary glass formed by cooling at 5 K/min to account for differences in thickness.

### D. Compressed exponential fitting to loss during annealing

The loss data for the $0.95T_g$ film shown in Figure 2 (red, lower panel) were normalized to the initial and final values and fitted to a KWW function, as shown in Supplemental Figure 7. The fitted value of $\tau_{KWW} = 3.47 \times 10^3$ s. The exponent $\beta = 1.1$ is great than 1, meaning that the process falls into the category of compressed relaxation.

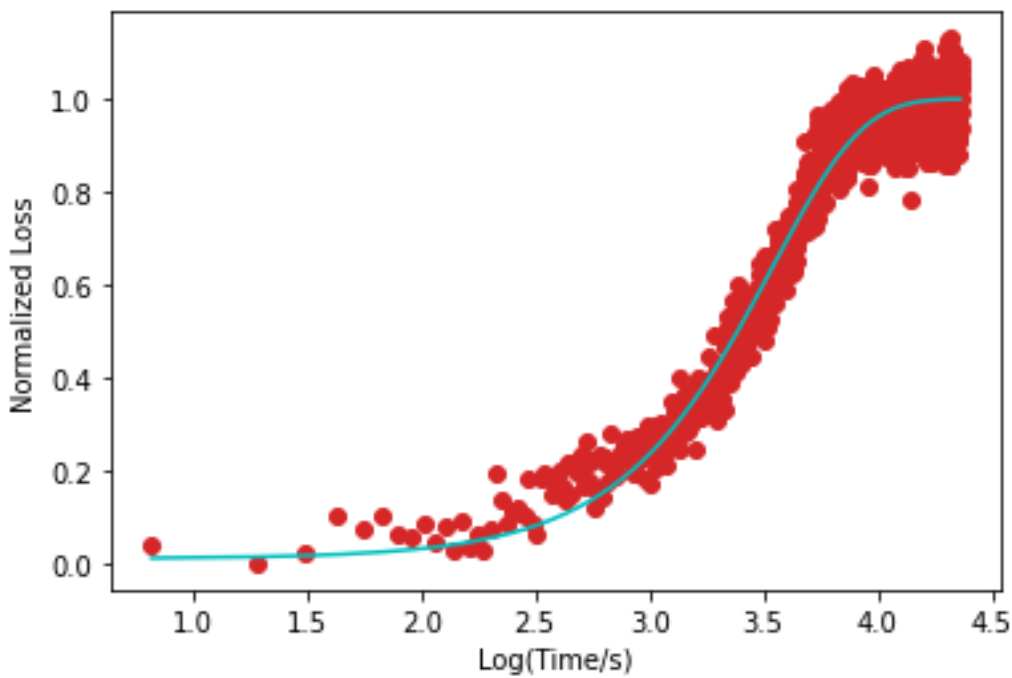

Supplemental Figure 7: Normalized loss during annealing of a $0.95T_g$ film at $0.98T_g$ for about 6 hours. The fitted KWW function is shown in cyan.

E. Second frequency during annealing

During the annealing of films of different stabilities at $0.98T_g$ for 6 hours, data at a second, higher, frequency was simultaneously collected. As seen in Supplemental Figure 8, this data reflects the same trends with annealing time as the 20 Hz data shown in Figure 2, with the $0.95T_g$ film rejuvenating completely and reaching a plateau at long times; the $0.90T_g$ film partially rejuvenating, reflected in smaller increases in both storage and loss; and the $0.85T_g$ and $0.80T_g$ films showing increases in the storage component but not the loss.

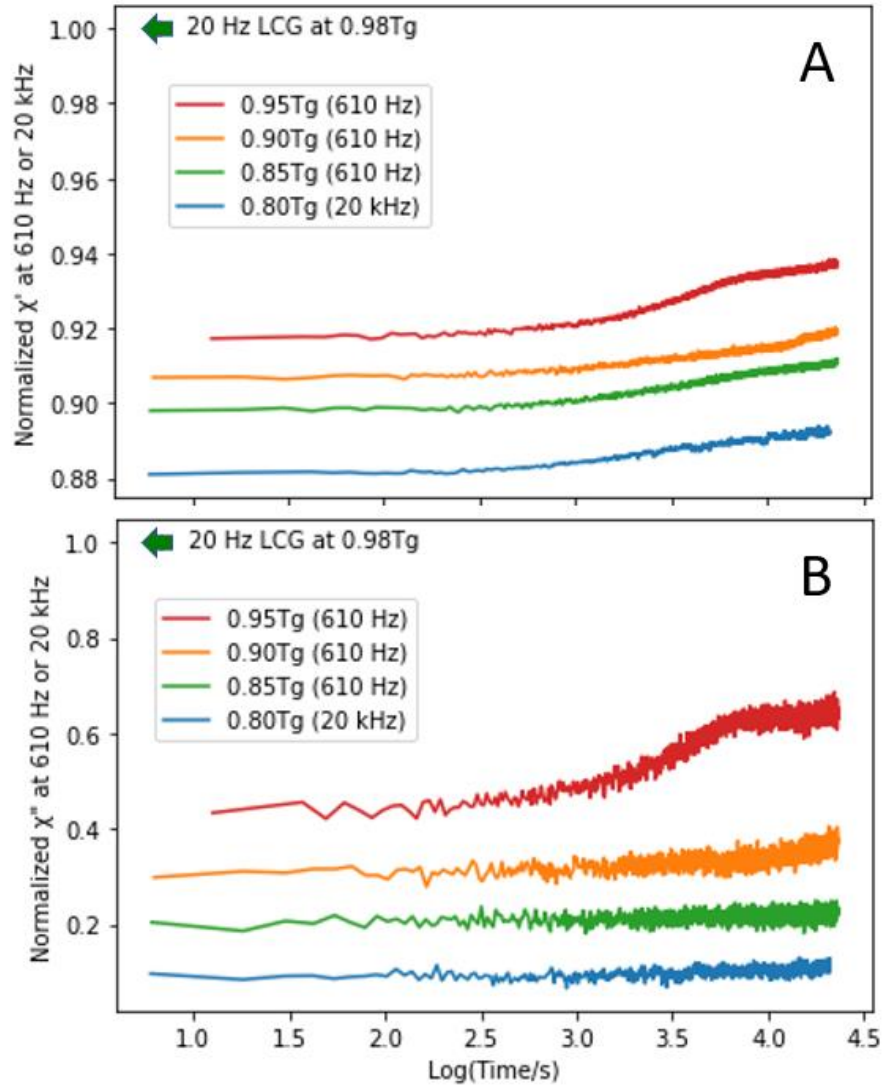

Supplemental Figure 8: Storage and loss of films deposited at the indicated deposition temperature during 6 hours of annealing at $0.98T_g$, measured at 610 Hz for $0.95T_g$, $0.90T_g$, and $0.85T_g$, and at 20 kHz for the $0.80T_g$ film. Equivalent to Figure 2 from the main text, recorded simultaneously during the same set of annealing experiments.

### F. Projected Transformation Times

Based on previous works, isothermal relaxation times correlate well with the relaxation time to a power of about 0.6, with a prefactor that depends on the stability of the film[61]. To establish this correlation for films used in this work, a series of films were deposited at $0.80T_g$ (to form the most stable glasses) and transformed isothermally at temperatures above $T_g$. If these transformation times are extrapolated to longer relaxation times (corresponding to lower annealing temperatures), they would be expected to transform on a timescale of almost 250 ks at $0.98T_g$, more than an order of magnitude longer than the annealing times used in this work.

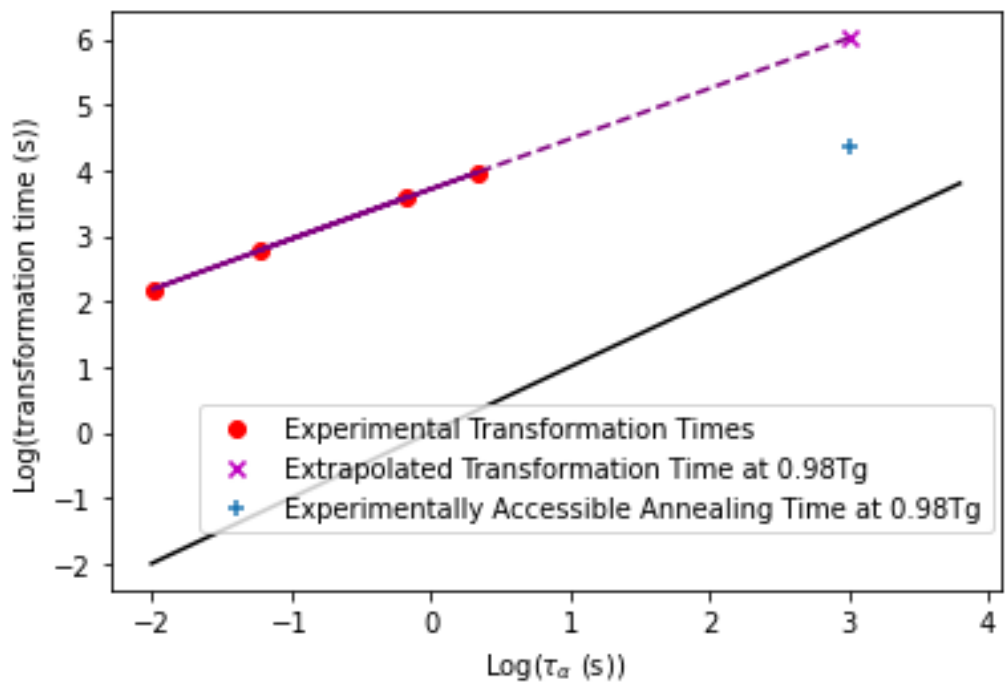

Supplemental Figure 9: Plot of isothermal transformation time as a function of α relaxation time on a log-log scale, with the line of 1:1 correspondence plotted in black. Experimental data points are shown as red circles, with the extrapolated transformation time at $0.98T_g$ is shown as the magenta 'x' along the purple line. The longest experimental annealing time is shown by the blue '+'.

### G. van Hove Function for Glass Particles.

We performed an additional test to check if the dynamics of the glass particles in the simulation evolved as the transformation of the entire system proceeded. Supplemental Figure 10 shows the van Hove function calculated for the glass particles at various waiting times, corresponding to different extents of transformation as indicated in the legend. The van Hove function was calculated at t = 1290, a value well into the plateau of the MSD (see Figure 8). Supplemental Figure 10 shows essentially identical van Hove functions, independent of the initial waiting time (and thus independent of the extent of transformation). This is consistent with Figure 8 and indicates that the properties of glass particles do not evolve as the total system begins to transform.

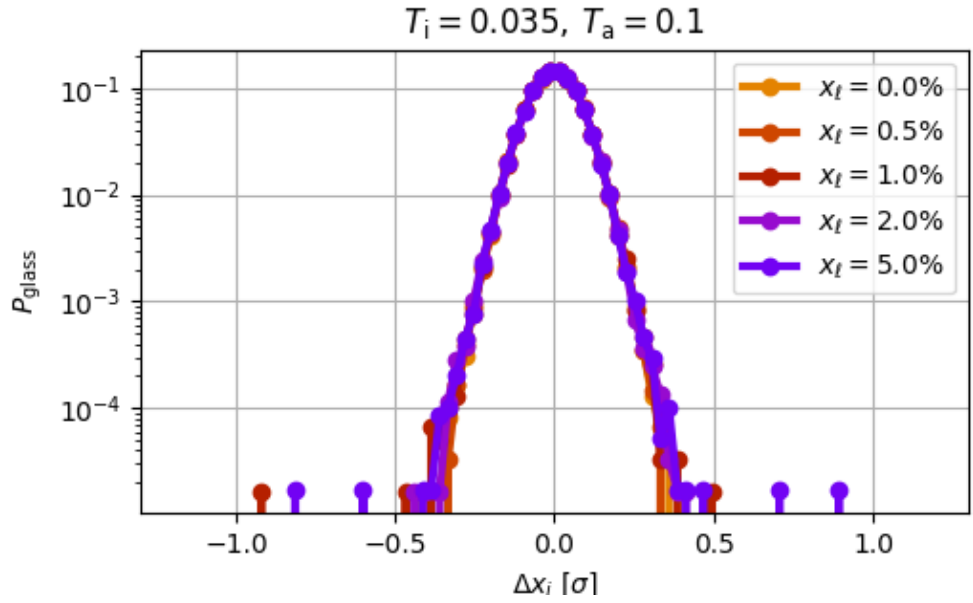

Figure 1: Van Hove function for the glass particles, calculated at t = 1290, with the calculation initiated at five different waiting times: $t_w/(10^6)$=0, 3.06, 3.67, 4.59 and 6.12. The waiting times correspond to liquid particle fractions of 0 – 5% as indicated in the legend.